\documentclass[twocolumn]{article} 

\usepackage{arxiv}
\usepackage[utf8]{inputenc} 
\usepackage[T1]{fontenc}    
\usepackage{xcolor}
\usepackage{hyperref}       
\usepackage{bbding}
\usepackage{caption}
\hypersetup{
    colorlinks = true,
    allbordercolors = white,
    urlcolor = black,
    citecolor = blue,
}
\usepackage{url}            
\usepackage{booktabs}       
\usepackage{amsfonts}       
\usepackage{nicefrac}       
\usepackage{microtype}      
\usepackage{lipsum}		
\usepackage{graphicx}
\usepackage{doi}
\usepackage{comment}
\usepackage{titlesec}
\usepackage{amsmath,amssymb}
\usepackage{enumitem}
\usepackage{capt-of}
\usepackage{url}
\usepackage{array}
\newcommand{\sgra}{Sgr A$^*$}
\newcommand{\uas}{\mu {\rm as}}
\newcommand{\solmass}{M_\odot}
\newcommand{\bhparams}{{\bf \Theta}}
\newcommand{\spin}{a}
\newcommand{\mass}{M}
\newcommand{\bgmag}{P_{\rm disk}}
\newcommand{\bgangle}{\xi_{\rm disk}}
\newcommand{\inc}{\theta_{\rm o}}
\newcommand{\velocity}{{\bf u}}
\newcommand{\momentum}{{\bf k}}
\newcommand{\magfield}{{\bf B}}
\newcommand{\pitch}{\phi_B}
\newcommand{\inclo}{\inc^{\rm lo}}
\newcommand{\inchi}{\inc^{\rm hi}}
\newcommand{\emission}{e}
\newcommand{\bfx}{{\bf x}}
\newcommand{\shear}{{\bf S}}
\newcommand{\rotation}{{\bf R}}
\newcommand{\angularvelocity}{\Omega}
\newcommand{\pixel}{{\bf p}}
\newcommand{\lc}{{\bf I}}

\newcommand{\measurements}{{\bf y}}
\newcommand{\netparams}{{\bf w}}
\newcommand{\mlp}{\operatorname{MLP}}

\newcommand{\jansky}{\text{Jy}}
\newcommand{\spectralindx}{\alpha_\nu}
\newcommand{\kepfrac}{f_{\rm K}}

\newcommand\blfootnote[1]{%
    \begingroup%
\let\thefootnote\relax\footnotetext{#1}%
\endgroup}%

\makeatletter
\renewcommand{\@makefntext}[1]{%
  \setlength{\parindent}{0pt}%
  \begin{list}{}{\setlength{\labelwidth}{1.5em}
    \setlength{\leftmargin}{\labelwidth}%
    \setlength{\labelsep}{3pt}%
    \setlength{\itemsep}{0pt}%
    \setlength{\parsep}{0pt}%
    \setlength{\topsep}{0pt}%
    \footnotesize}%
  \item[\@thefnmark\hfil]#1
  \end{list}%
}
\renewcommand\@makefntext[1]{%
\setlength\parindent{1em}%
\noindent
\mbox{\@thefnmark}{#1}}
\makeatother

\newcommand{\apj}{Astrophys. J.}   
\newcommand{\apjl}{Astrophys. J. Lett.}   
\newcommand{\aap}{Astron. Astrophys.}   
\newcommand{\mnras}{Mon. Not. R. Astron. Soc.}   
\newcommand{\prd}{Phys. Rev. D}   

\title{Orbital Polarimetric Tomography of a Flare Near the Sagittarius A$^*$ Supermassive Black Hole}

\author{
Aviad Levis\href{https://orcid.org/0000-0001-7307-632X}{\includegraphics[scale=0.06]{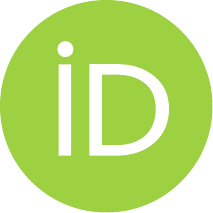}}$^{1*}$,
Andrew A.~Chael\href{https://orcid.org/0000-0003-2966-6220}{\includegraphics[scale=0.06]{figures/orcid.pdf}}$^2$, 
Katherine L.~Bouman\href{https://orcid.org/0000-0003-0077-4367}{\includegraphics[scale=0.06]{figures/orcid.pdf}}$^1$, 
Maciek Wielgus\href{https://orcid.org/0000-0002-8635-4242}{\includegraphics[scale=0.06]{figures/orcid.pdf}}$^3$, 
Pratul P.~Srinivasan\href{https://orcid.org/0009-0008-8268-3285}{\includegraphics[scale=0.06]{figures/orcid.pdf}}$^4$ \vspace{0.2cm} \\ 
$^1$California Institute of Technology, Pasadena, CA, USA. \\
$^2$Princeton Gravity Initiative, Princeton University, Princeton, NJ, USA. \\
$^3$Max-Planck-Institut für Radioastronomie, Bonn, Germany. \\
$^4$Google Research, San Francisco, CA, USA. \\
$^{*}${\tt aviad.levis@gmail.com}
}
\date{}

\titlespacing*{\section}{0pt}{0pt}{0pt}
\titlespacing*{\subsection}{0pt}{0pt}{0pt}
\titlespacing*{\subsubsection}{0pt}{0pt}{0pt}
\titlespacing*{\abstract}{0pt}{0pt}{0pt}

\begin{document}
\setlength{\abovedisplayskip}{6.5pt}
\setlength{\belowdisplayskip}{6.5pt}

\twocolumn[
  \begin{@twocolumnfalse}
    \maketitle
    \vspace{-0.75cm}
    \begin{abstract}
    The interaction between the supermassive black hole at the center of the Milky Way, Sagittarius A$^*$, and its accretion disk occasionally produces high-energy flares seen in X-ray, infrared, and radio. One proposed mechanism that produces flares is the formation of compact, bright regions that appear within the accretion disk and close to the event horizon. Understanding these flares provides a window into accretion processes. Although sophisticated simulations predict the formation of these flares, their structure has yet to be recovered by observations. Here we show the first three-dimensional (3D) reconstruction of an emission flare recovered from ALMA light curves observed on April 11, 2017. Our recovery shows compact, bright regions at a distance of roughly six times the event horizon. Moreover, it suggests a clockwise rotation in a low-inclination orbital plane, consistent with prior studies by GRAVITY and EHT. To recover this emission structure, we solve an ill-posed tomography problem by integrating a neural 3D representation with a gravitational model for black holes. Although the recovery is subject to, and sometimes sensitive to, the model assumptions, under physically motivated choices, our results are stable, and our approach is successful on simulated data.
    \end{abstract}
    \vspace{0.5cm}
  \end{@twocolumnfalse}
]

\begin{figure*}[t]
	\centering \includegraphics[width=\linewidth]{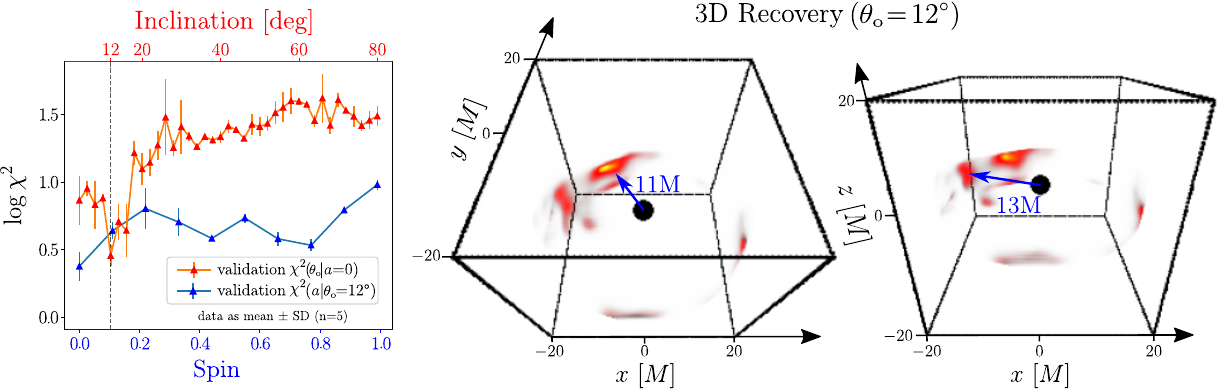}
	\caption{A 3D recovery of a \sgra{} flare observed by ALMA on April 11, 2017. [Left panel] The {\em validation-$\chi^2$}, a robust data-fitting metric (see Methods section), indicates a preference of low inclination angles, $\inc < 18^\circ$, with a local minimum around $\inc = 12^\circ$ (red curve). For each inclination, the 3D recovery is run with five random initialization producing a spread that indicates recovery stability. The blue curve indicates that the analysis is largely insensitive to the black hole spin. [Right panels] A recovered 3D volume visualized from two view angles in intrinsic (flat space) coordinates (the event horizon illustrated for size comparison). The recovery shows two emission regions (blue arrows) at radii of $11 - 13~{\rm M}$ ({$\sim 6$} times the Schwarzschild radius).}
	\label{fig:alma_rec}
\end{figure*}

The compact region around the Galactic Center supermassive black hole \sgra{} is a unique environment where the magnetized turbulent flow of an accretion disk is subject to extreme gravitational physics. The dynamical evolution of this complex system occasionally leads to the production of energetic flares~\cite{genzel2003near} seen in X-ray~\cite{neilsen2013chandra}, infra-red~\cite{fazio2018multiwavelength}, and radio~\cite{Wielgus2022lc}. The physical nature, structure, origin, formation, and eventual dissipation of flares are topics of active research~\cite{fazio2018multiwavelength,Marrone_2008,Haggardetal2019,dexter2020sgr,Witzel_2021} key to our understanding of accretion flows around black holes.
One proposed explanation for \sgra{} flares is the formation of compact bright regions caused by hot pockets of lower-density plasma within the accretion disk, that are rapidly energized (e.g. through magnetic reconnection~\cite{broderick2005imaging}). These ``bubbles'', ``hotspots'' or ``flux tubes'', observed in numerical simulations (e.g. \cite{ripperda2022black}), are hypothesized to form in orbit close to the innermost stable circular orbit (ISCO) of \sgra{}. The association of flares with orbiting hotspots close to the event horizon is consistent with near-infrared detections made by the GRAVITY Collaboration~\cite{gravity18, gravity20} and radio observations of the Atacama Large Millimeter/Submillimeter Array (ALMA)~\cite{wielgus2022orbital}.

The context for this work is set by the first images~\cite{akiyama2022firstI} of \sgra{} revealed by the Event Horizon Telescope (EHT) collaboration. The images, reconstructed from Very Long Baseline Interferometry (VLBI) observations from April 6–7, 2017, show a ring-like structure with a central brightness depression -- a strong suggestion that the source is indeed a supermassive black hole~\cite{akiyama2022firstVI}. Even in its quiescent state imaged by EHT on April 6/7, \sgra{} has shown significant structural variability~\cite{akiyama2022firstIII}. On April 11, 2017, \sgra{} was observed by ALMA directly after a high-energy flare seen in X-ray. The ALMA light curves exhibit an even higher degree of variability to April 6/7~\cite{Wielgus2022lc, akiyama2022firstII}, including distinct coherent patterns in the linear polarization~\cite{wielgus2022orbital} with variability on the scale of an orbit. The presence of synchrotron-radiating matter very close to the horizon of \sgra{} could potentially give rise to bright 3D structures that orbit and evolve within the accretion disk. In this work, we present the first 3D recovery of emission in orbit around \sgra{} reconstructed from ALMA light curves observed on April 11, 2017 (Fig.~\ref{fig:alma_rec}).

To achieve this 3D reconstruction we develop a novel computational approach which we term: {\em orbital polarimetric tomography}. In contrast to prior work by \cite{gravity18} and \cite{wielgus2022orbital}, which employed a strongly constrained parametric hotspot model, with only a handful of parameters to tune and interpret the observations, the goal of this work is to recover the complex 3D structure of flares as they orbit and evolve in the accretion disk around \sgra{}. 

Tackling this inverse problem necessitates a change from typical tomography, wherein 3D recovery is enabled by multiple viewpoints. Instead, the tomography setting we propose relies on observing a structure in orbit, traveling through curved spacetime, from a fixed viewpoint. As it orbits the black hole, the emission structure is observed (projected) along different curved ray paths. These observations of the evolving structure over time effectively replace the observations from multiple viewpoints required in traditional tomography. Our approach builds upon prior work on dynamical imaging and 3D tomography in curved spacetime, which showed promising results in {\em simulated} future Event Horizon Telescope (EHT) observations~\cite{Tiede_2020, levis2021inference, levis2022gravitationally}. 

Similar to the computational images recovered by EHT~\cite{akiyama2022firstIII}, our approach solves an under-constrained inverse problem to fit a model to the data. Nevertheless, ALMA observations do not resolve event horizon scales ($\sim 10^5$ lower resolution), which makes the tomography problem we propose particularly challenging. To put it differently, we seek to recover an evolving 3D structure from a single-pixel observation over time. In order to solve this challenging task, we integrate the emerging approach of neural 3D representations~\cite{levis2022gravitationally, mildenhall2020nerf}, which has an implicit regularization that favors smooth structures~\cite{tancik2020fourfeat}, with physics constraints (details in the Methods section). The robustness of the results thus relies on the validity of the constraints imposed by the gravitational and synchrotron emission models.

We take advantage of the very high signal-to-noise and cadence of the ALMA dataset~\cite{Wielgus2022lc}, as well as the linear polarization information~\cite{wielgus2022orbital}. The choice to only fit the LP light curves reflects the uncertainty associated with the un-polarized intensity of the background accretion disk. While the total intensity light curve is dominated by the accretion disk, such extended emission structures are partially depolarized in an image-average sense~\cite{wielgus2022orbital}. In contrast, compact bright sources, such as a putative hotspot, are characterized by a large fractional linear polarization (LP) and fast evolution on dynamical timescales~\cite{wielgus2022orbital, Gravity2023}, hence allowing separation of the flare component from the background accretion. In the Supplementary (Sec.~2.2), we quantitatively assess the effect of the background accretion disk on simulated reconstruction results.

\begin{figure*}[t]
	\centering \includegraphics[width=0.9\linewidth]{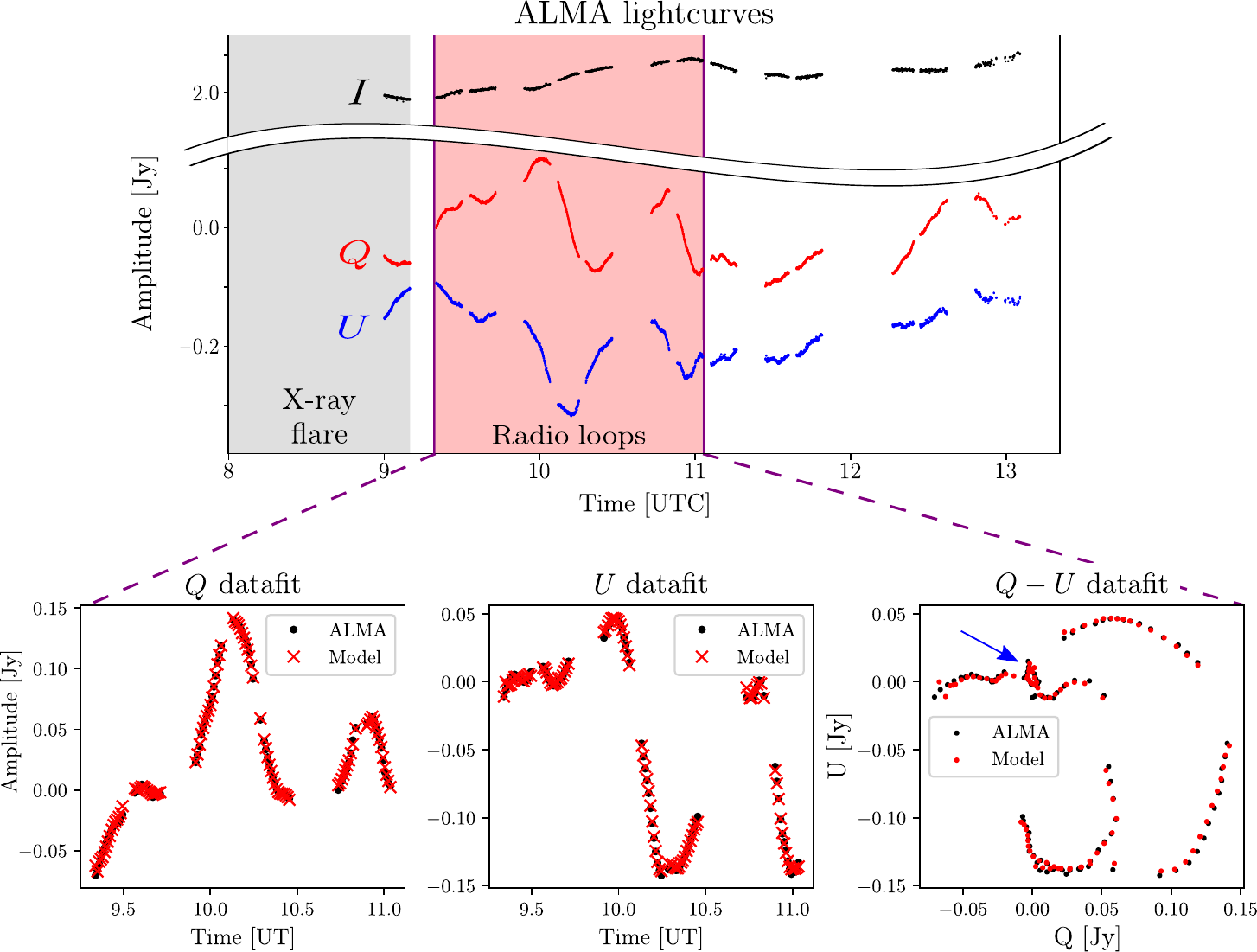}
	\caption{ALMA light curves and a model fit over the a period of $\sim100$ minutes. [Top] The 229 GHz light curves were observed on 2017 April 11 (MJD 57854) as part of the EHT \sgra{} campaign. The red-shaded region corresponds to a time of period $\sim 100$ minutes in which polarimetric (Q-U) loops are apparent, directly after an X-ray flare was observed (gray-shaded region). The rotation of the polarization angle at a period similar to a Keplerian orbital period suggests the signal is coming from a bright compact structure in orbit around \sgra{}~\cite{wielgus2022orbital}. [Bottom] A data-fit is preformed on the {\em intrinsic} LP curves (centered and de-rotated). The model light curves are produced through ray tracing the estimated 3D volume at a fiducial inclination angle of $\inc=12^\circ$. The resulting model light curves accurately describe the data including the small looping feature highlighted by the blue arrow.}
	\label{fig:alma_datafit}
\end{figure*}

\section*{Results}
On April 11, 2017, ALMA observed \sgra{} at $\sim 230 {\rm GHz}$ as part of a larger EHT campaign (Fig.~\ref{fig:alma_datafit} top). The radio observations directly followed a flare seen in the X-ray. The LP, measured by ALMA-only light curves~\cite{Wielgus2022lc,wielgus2022orbital} as a complex time series $Q(t)+iU(t)$, appears to evolve in a structured, periodic, manner suggesting a compact emission structure in orbit. The work of \cite{wielgus2022orbital} hypothesizes a simple bright spot (i.e. idealized point-source~\cite{gelles2021polarized} or spherical Gaussian~\cite{vos2022polarimetric}) at \mbox{$r\sim 11 {\rm M}$}, however, a rigorous data-fitting was not performed. Furthermore, the proposed parametric model is limited and does not explain all of the data features. The orbital polarimetric tomography approach that we propose enables a rigorous data-fitting and recovery of flexible 3D distributions of the emitting matter, relaxing the assumption of a fixed orbiting feature enforced by prior studies~\cite{gravity18, wielgus2022orbital, yfantis2023}. This opens up a new window into understanding the spatial structure and location of flares relative to the event horizon.

Our model, detailed in the Methods section, is able to fit the ALMA light curve data very accurately (Fig.~\ref{fig:alma_datafit} bottom). The optimization procedure simultaneously constrains the inclination angle of the observer and estimates a 3D distribution of the emitting matter associated with this flaring event, starting from 9:20 UT ($\sim 30$ minutes after the peak of the X-ray flare~\cite{wielgus2022orbital}). Despite the fact that ALMA observations are unresolved (effectively a single pixel with time-dependent complex LP information) at the horizon scale, our analysis suggests some interesting insights:
\setlist{nolistsep}
\begin{itemize}[leftmargin=*, topsep=0pt, itemsep=0.5em]
    \item Low inclination angles ($\inc<18^\circ$, Fig.~\ref{fig:alma_rec} left panel, red) are preferred by the {\em validation-$\chi^2$} (Methods section). While the methodology is different, this result is broadly consistent with EHT findings from April 6/7~\cite{akiyama2022firstV}, which favored low inclination angles of $\sim 30^\circ$ by comparing recovered images with General Relativistic Magneto Hydro Dynamic simulations. The fiducial model of \cite{wielgus2022orbital} corresponded to an inclination angle of $\sim 22^\circ$. Low inclination was also favored in the analysis of the GRAVITY infra-red flares~\cite{gravity18, gravity20,Gravity2023}.
    \item The recovered 3D emission has two compact bright regions at \mbox{$r \sim 11M, 13M$} (Fig.~\ref{fig:alma_rec} middle/right panels). The location (radius and azimuthal position) of the bright region is consistent with the qualitative analysis of \cite{wielgus2022orbital, yfantis2023}.
\end{itemize}

\begin{figure*}[t]
	\centering \includegraphics[width=\linewidth]{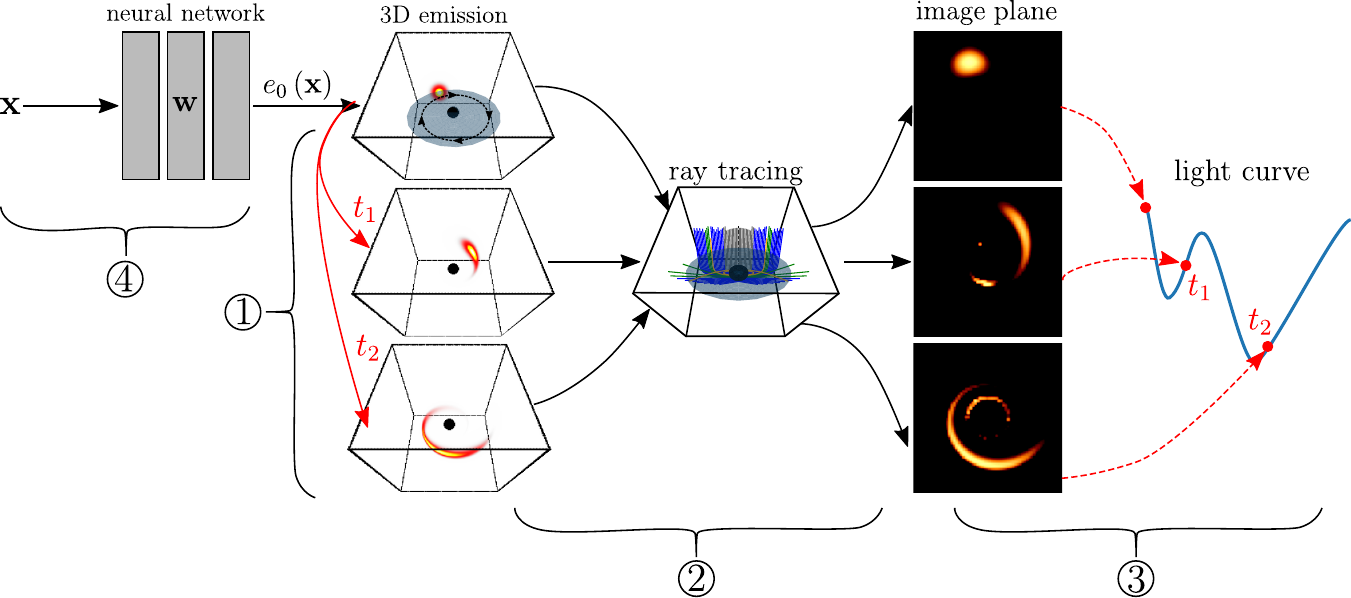}
	\caption{An overview of the orbital tomography framework based on light curve observations. 1. An orbital model propagates an initial (canonical) emission volume ($\emission_0$) in time.  2. Ray tracing: we compute General Relativistic (GR) ray paths according to the black hole parameters and numerically integrate the 3D emissivities to synthesize image plane frames. 3. Each frame is summed to produce a single light curve data point which downstream is compared to the observations. 4. A neural representation of the underlying 3D volume. Each component is extensively discussed in the Methods section.}
	\label{fig:overview}
\end{figure*}

\subsection*{Data fitting}
Before solving the tomography problem, we perform pre-processing according to the procedure outlined in \cite{wielgus2022orbital}. In particular, we time average the data, subtract a constant (time-averaged) LP component interpreted as the ring-like accretion disk component observed by the EHT, and de-rotate the electric vector polarization angle (EVPA) to account for Faraday rotation (details in the Methods section). Figure \ref{fig:alma_datafit} illustrates the data before and after the pre-processing. 

To obtain a model prediction a 3D emission structure is adjusted so that, when placed in orbit, the numerically ray-traced light curves align with the observations. To recover the vertical structure, our approach primarily leverages asymmetries in the polarimetric radiative transfer. In particular, the geometry of space-time and the magnetic field dictate the angle of linear polarization (Q-U). Moving an emission point changes the observed angle of linear polarization. Thus, erroneously placing emission at $t=0$ and propagating it in time will rotate to the overall linear polarization in directions that are incompatible with the observed Q-U time series. 

Computing the model predictions relies on ray tracing, which requires knowledge of the path rays take in 3D curved spacetime. These geodesic paths depend on the {\em unknown} black hole properties~\cite{gelles2021polarized}: mass, spin, and inclination. Nevertheless, the mass of \sgra{} is well constrained through stellar dynamics~\cite{ghez2008measuring}; \mbox{$\mass\simeq 4 {\times} 10^6 \solmass$} where $\solmass$ denotes solar mass. Furthermore, Fig.~\ref{fig:alma_rec} (blue curve) illustrates that the data-fit is not very sensitive to black hole spin: $\spin\in[0,1]$. Thus, the only remaining unknown is the inclination angle. 

To estimate the inclination, we numerically bin $\inc \in [0,\pi/2]$ and recover the 3D emission for every given (fixed) angle. For each angle, we recover a (locally) optimal 3D emission by minimizing a $\chi^2$ loss over the model parameters. Practically, for numerical stability, we avoid the extreme angles of face-on and edge-on by gridding $\inc\in[4^\circ, 80^\circ]$ (at $2^\circ$ increments). Figure \ref{fig:alma_rec} plots the validation-$\chi^2$: a robust likelihood approximation for $\inc$, which appears to favor low inclination angles (details in the Methods section). For each inclination, the recovery is run five times with a random initialization for the 3D structure. Therefore, the error bars are not a measure of posterior uncertainty; rather, they indicate the stability of the locally optimal solution.

An overview of the tomographic data-fitting framework is illustrated in Fig.~\ref{fig:overview}. Mathematically, the 3D emission volume is estimated by minimizing a $\chi^2$ loss between the observed LP and the model prediction. The continuous 3D emission volume is represented by a coordinate-based, fully-connected, neural network (``neural representation'') and is constrained to a domain with a radius of $6{\rm M} \leq r \leq 20 {\rm M}$ and close to the equatorial disk $|z| \leq 4{\rm M}$ ($6M$ is the innermost stable circular orbit of a non-spinning black hole). The data-fit  used in this work relies on the reduced $\chi^2$ definitions of \cite{akiyama2019firstIV}. This is not a strict definition of reduced $\chi^2$ that includes degrees of freedom in normalization. Rather, it is normalized by the total number of data points, which is useful for comparing fit quality in our experiments where degrees of freedom remain fixed.
\begin{figure*}[t]
	\centering \includegraphics[width=\linewidth]{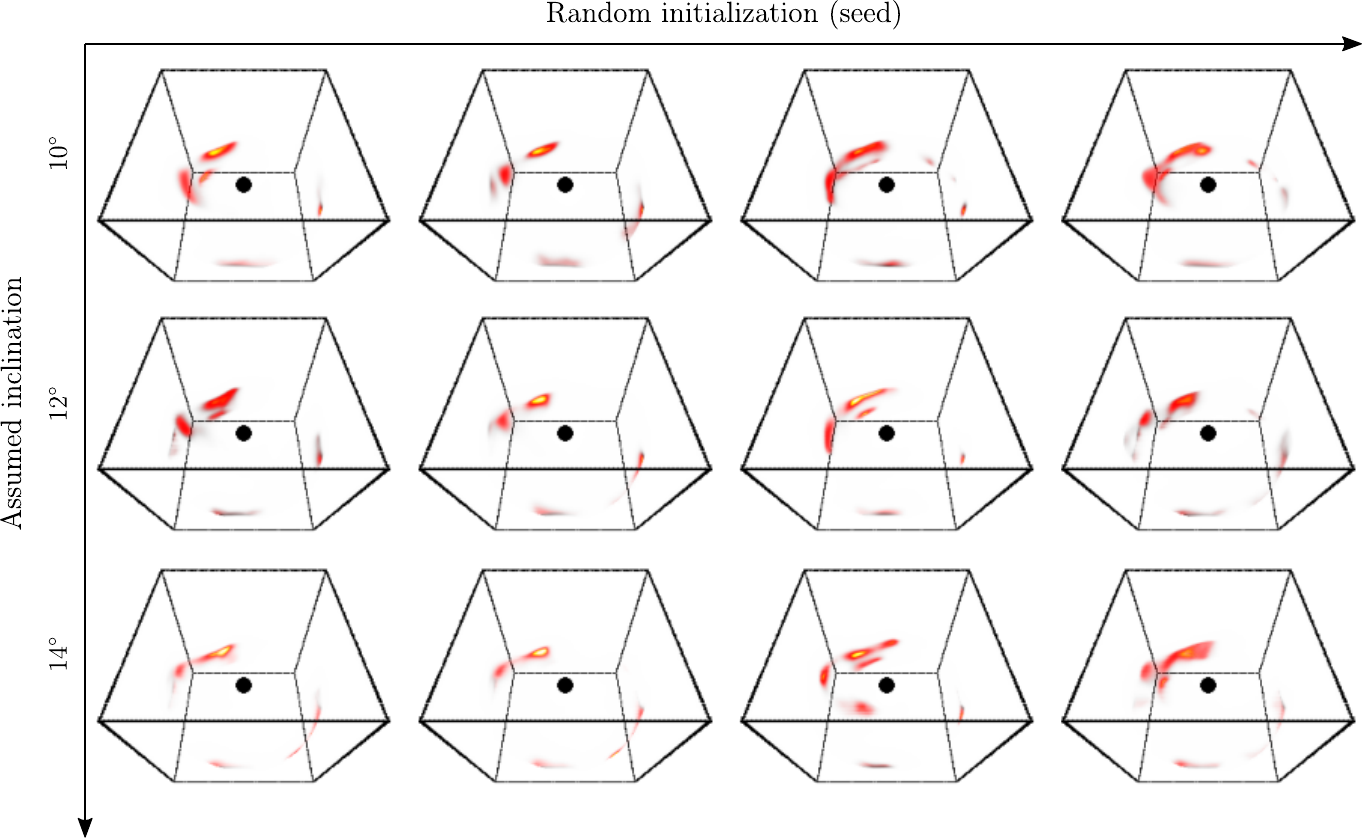}
	\caption{A visualization of 3D recoveries across different inclinations and initial conditions. While some of the details of the recovered structure depend on both axes, some key features remain consistent. The exact angular extent of the structures is not stable, nevertheless, the azimuthal and radial position appears stable and consistent with the average structure highlighted in Fig.~\ref{fig:alma_rec}. Moreover, the separation into distinct emission regions of an elongated feature trailed by a smaller, dimmer, compact bright spot appears consistent across the different recoveries.}
	\label{fig:alma_3d_variability}
\end{figure*}

The ill-posed inverse problem we solve does not have a unique solution. The recovered 3D structure depends, among other factors, on the assumed inclination angle. Furthermore, solving a non-convex optimization problem with stochastic gradient descent methods leads to a local (and not global) minimum. Thus, the recovered 3D structure also depends on the random initialization of the network weights. Figure \ref{fig:alma_3d_variability} highlights the robustness of the recovered 3D structure across different inclinations and initial conditions. While this is not an exploration of the posterior distribution, the different recoveries give a sense of the solution's stability. Qualitatively, the details of each recovered structure exhibit dependence on both the inclination angle and initialization. Nevertheless, some key features are consistent across these two axes. While the exact angular extent of the structures is not stable, the azimuthal and radial positions appear stable and consistent with the average structure. Moreover, the separation of the emission into two distinct structures appears consistent across the different recoveries. 

\begin{figure}[t]
	\centering \includegraphics[width=\linewidth]{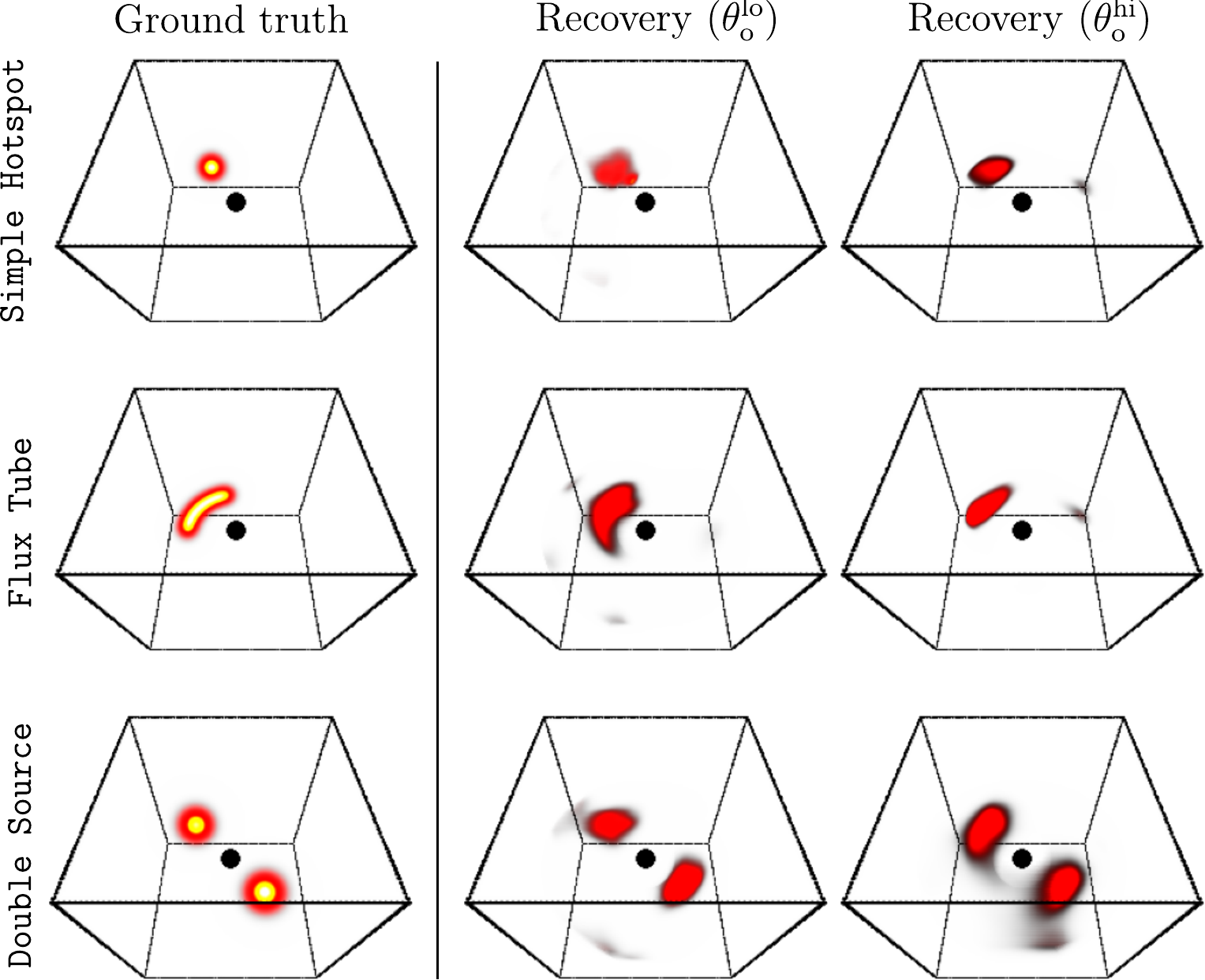}
	\caption{3D recoveries for three simulated structures observed at two (unknown) inclination angles: $\inclo=12^\circ$, $\inchi=64^\circ$. Using synthetically generated light-curves as observations, the 3D reconstructions are able to recover different flare morphologies in the presence of background accretion noise (not visualized in this figure). Further analysis and details are given in the Supplementary (Sec.~2.3).}
	\label{fig:sim_recoveries}
\end{figure}

To analyse the ability to recover and detect different underlying 3D morphologies, we simulated synthetic datasets mimicking ALMA observations for three underlying 3D structures: {\tt Simple Hotspot},  {\tt Flux Tube}, {\tt Double Source}. Figure \ref{fig:sim_recoveries} highlights the recovery results obtained from these datasets at two (unknown) inclination angles. A comprehensive analysis of the simulated datasets and reconstruction results is given in the Supplementary (Sec.~2).

\subsection*{Assumptions and systematic noise}
\begin{figure*}[t]
	\centering \includegraphics[width=.95\linewidth]{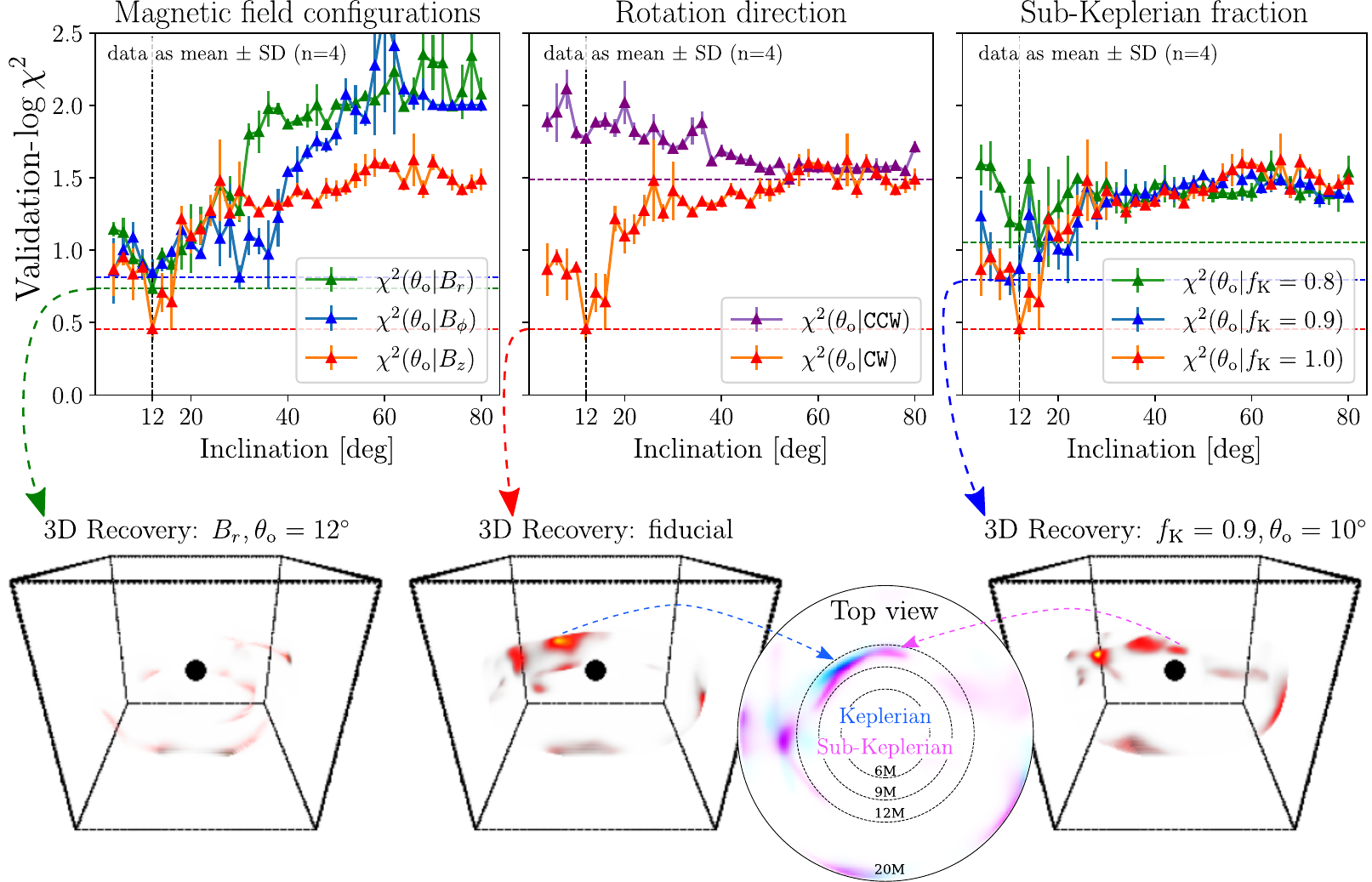}
	\caption{The effects of different model choices: magnetic field, rotation direction and orbital velocity. [Top] Validation $\chi^2$ under each model choice. [Bottom] 3D reconstructions under various model assumptions. The red curve in all panels represents the fiducial model parameters: vertical magnetic field ($B_z$), clockwise rotation ({\tt CW}), and Keplerian orbit ($\kepfrac=1.0$). The global minimum for each curve is highlighted by a horizontal dashed line in the respective color. [Left] Three different magnetic field configurations: vertical, radial, and toroidal (subscripts: $z, r, \phi$ respectively). The recovered 3D under a radial magnetic field appears spread out rather than a compact hot-spot like structure. That being said, consistent with the analysis of \cite{gravity18, wielgus2022orbital}, vertical magnetic fields, which do result in a compact hotspot-like structure, are favorable according to this metric with lower validation-$\chi^2$ around $\inc=12^\circ$. 
    [Center] A comparison of clockwise ({\tt CW}) and counter-clockwise ({\tt CCW}) angular velocity models. Consistent with the analysis of \cite{wielgus2022orbital}, a {\tt CW} rotation is preferred across all inclination angles. [Right] A Keplerian orbit has the lowest validation-$\chi^2$ fit across three different fractions of sub-Keplerian orbit: $\kepfrac=1.0,0.9,0.8$. The recovery under $\kepfrac=0.9$ (bottom right) is broadly consistent with the Keplerian model with a tendency towards smaller radii (illustrated by the top view panel).}
	\label{fig:magnetic_fits_cw_ccw}
\end{figure*}

Recovering the 3D structure from light curve observations is highly ill-posed. Thus, the recovery relies on physical constraints and model choices that we impose through the gravitational and synchrotron emission models. The robustness of the results depends on the validity of these model choices, detailed in Table \ref{table:1}, and discussed below.

The key assumption for orbital tomography is that the emission is in orbit within an accretion disk, near the equatorial plane, and can be modeled as a simple transformation to a canonical (or initial) 3D emission. Note that small shifts from the equatorial plane are allowed by our model. This enables the formulation of an inverse problem for estimating the 3D emission from observations. We consider orbits characterized by a Keplerian angular velocity profile (neglecting radial or vertical velocity components), accounting for shearing due to differential rotation (ignored by the previous analyses~\cite{gravity18, wielgus2022orbital, yfantis2023}) while neglecting the dynamics of cooling, heating, expansion, and turbulence. While this simplifying assumption does not hold in general, it is consistent with theoretical simulations~\cite{ripperda2022black}, which show consistent structures on short $\sim 1$ orbit timescales.

Furthermore, in modeling synchrotron emission, we assume a homogeneous vertical magnetic field that is externally fixed and is independent of the flare or accretion disk dynamics. The choice of a vertical magnetic field for the fiducial recovery is motivated by the notion that vertical magnetic fields could be powering \sgra{} flares, apparent in GRMHD simulations that produce magnetic eruption events~\cite{ripperda2022black}. Moreover, from an observational standpoint, vertical magnetic fields are preferred by both the near infrared analysis of GRAVITY~\cite{gravity18,gravity20,Gravity2023} and mm ALMA analysis of \cite{wielgus2022orbital}. Nevertheless, the true spatial structure and dynamic properties of the magnetic fields around \sgra{} are largely unknown.

In Fig.~\ref{fig:magnetic_fits_cw_ccw} we analyze some of the systematic model choices detailed above by exploring the effects of a) magnetic field configuration b) rotation direction and c) sub-Keplerian orbits on the data fit and 3D reconstruction. It is important to note that we do not aim to exhaustively test all possible magnetic field and orbital velocity models, but instead highlight the sensitivity of our reconstruction to these model choices. The top-left panel of Fig.~\ref{fig:magnetic_fits_cw_ccw} compares the validation-$\chi^2$ for three magnetic field configurations: vertical, radial, and toroidal, respectively denoted by subscripts: $z, r, \phi$. For a radial magnetic field, the best fit recovery is not a compact bright emission region (Fig.~\ref{fig:magnetic_fits_cw_ccw} bottom left). Rather, it is a fainter, diffuse structure. Even so, according to the data-fit, and consistent with prior studies, vertical magnetic fields are preferred with a lower validation-$\chi^2$ value. 

The center and rightmost panels in Fig.~\ref{fig:magnetic_fits_cw_ccw} highlight how clockwise rotation ({\tt CW}) and a Keplerian orbit are favorable to counter-clockwise rotation ({\tt CCW}) or sub-Keplerian orbits, consistent with the analyses of GRAVITY~\cite{Gravity2023}, and \cite{wielgus2022orbital}. To test the fit of orbit direction and sub-Keplerian fraction, we set the angular velocity profile to $\angularvelocity = \pm \kepfrac \angularvelocity_{\rm K}$ where the $\pm$ sign dictates the direction ({\tt CW/CCW}) and $\kepfrac$ the magnitude ($\kepfrac = 1$ results in a clockwise Keplerian orbit). The 3D recovery under $\kepfrac = 0.9$ is shown in the bottom-right panel of Fig.~\ref{fig:magnetic_fits_cw_ccw}, highlighting a broadly consistent recovery with the fiducial model assumptions. However, this consistency eventually breaks down at strong deviations from the Keplarian velocity assumption (also resulting in lower validation-$\chi^2$ values). The top view illustrates how a sub-Keplerian orbit impacts the recovered flare's radial position.

Another key assumption made by our work is that the millimeter emission region is optically thin. This is consistent with both EHT observations of \sgra{}~\cite{akiyama2022firstII} and their theoretical interpretation~\cite{akiyama2022firstIII, akiyama2022firstV}. Moreover, theoretical analysis~\cite{ripperda2022black} has shown that a ``flux tube'' flare would be optically thinner due to its higher temperature and lower density compared to the surrounding accretion flow.

Finally, we assume that the recovered emission structure is within the accretion disk. While our model does not account for an alternative jet interpretation, the assumption of accretion flares is consistent with a) theoretical simulations showing powerful equatorial current sheaths associated with flux eruptions forming within the accretion disk~\cite{ripperda2022black}; b) observational evidence from EHT / VLBI analyses consistent with a compact source model without any detectable jet contribution~\cite{akiyama2022firstII, Issaoun2019, cho2022} and c) the GRAVITY detection, indicating an astrometric center aligning with the mass center, implies that the orbiting feature is in proximity to the equatorial plane. The alternative scenario demands a precise alignment between the direction of the jet and the observer's line of sight.

\section*{Discussion}
We present a novel computational approach to image dynamic 3D structures orbiting the most massive objects in the universe. Integrating polarimetric general relativistic ray tracing and neural radiance fields enables resolving a highly ill-posed tomography in the extremely curved space-time induced by black holes. Applying this approach to ALMA observations of \sgra{} reveals a 3D structure of a flare, with a location broadly consistent with the qualitative analysis presented in \cite{wielgus2022orbital}. This first attempt at a 3D reconstruction of a \sgra{} flare suggests an azimuthally elongated bright structure at a distance of $~11{\rm M}$ trailed by a dimmer source at $~13 {\rm M}$. Although the recovered 3D is subject to, and sometimes sensitive, to the gravitational and emission models, under physically motivated choices, we find that the 3D reconstructions are stable and our approach is successful on simulated data. Moreover, our data-fit metrics provide constraints favoring low inclination angles and clockwise rotation of the orbital plane, supporting the analyses of \cite{wielgus2022orbital}, EHT~\cite{akiyama2022firstIII}, and GRAVITY~\cite{gravity18}. 

Orbital polarimetric tomography shows great promise for 3D reconstructions of the dynamic environment around a black hole. Excitingly, extending the approach and analysis to spatially resolved observations (e.g., EHT) and multi-frequency data could enable relaxing assumptions to further constrain the underlying physical structures that govern the black hole and plasma dynamics (e.g., black hole spin, orbit dynamics, magnetic fields). To that end, future work will likely need to extend our model to non-optically thin media and non-azimuthal velocity patterns. Lastly, by adapting orbital polarimetric tomography to other rich sources of black hole time series observations (e.g., quasars and microquasars), this imaging technology could open the door to population statistics and improve our understanding of black holes and their accretion processes.
\renewcommand{\arraystretch}{1.2}
\begin{table*}[t]
\centering
\begin{tabular}{ |l | c | } 
\hline
 Emission model & \begin{tabular}{@{}c@{}}
                    Synchrotron\\
                    fixed vertical magnetic field;~~~ optically thin disk\\
                 \end{tabular} \\
 \hline
 Dynamical model & \begin{tabular}{@{}c@{}}
                    Keplerian\\
                    $t_0=$9:20 UT;~~~clockwise orbit (no radial / vertical velocity);~~~ velocity shear \\
                 \end{tabular} \\
 \hline
 Gravitational model & \begin{tabular}{@{}c@{}}
                    Kerr\\
                    mass $= 4.154{\times}10^6 \solmass$;~~~ non spinning;~~~ $\inc$ estimated from data\\
                 \end{tabular} \\
 \hline
  3D model & \begin{tabular}{@{}c@{}}
                    Neural representation\\
                    Recovery domain: $6{\rm M} \leq r \leq 20 {\rm M}$ (FOV $\approx 200 \uas$);~~~ $|z| \leq 4{\rm M}$ \\
                 \end{tabular} \\
\hline
\end{tabular}
\vspace{0.1cm}
\caption{Summary of the key physical assumptions made in the modelling. We assume that the emission source is in orbit around a black hole within its accretion disk. The recovered 3D emission relies on combining the flexibility of 3D neural fields with black-hole physics. Thus, the accuracy of the reconstruction depends on the validity of the model assumptions. Fig.~\ref{fig:magnetic_fits_cw_ccw} explores the effects of some of the assumptions (magnetic field configurations, orbit direction, and sub-Keplerian orbits) on both the data-fit and recovered 3D. We assume a non-spinning black hole since our analysis found that results are only weakly sensitive to black hole spin (Fig.~\ref{fig:alma_rec} left panel)}
\label{table:1}
\end{table*}

\section*{Methods}
In the following section, we describe our methodology, which is evaluated on synthetic simulations and analyzed in the Supplementary Material.

\subsection*{Pre-processing}
We reduced the ${\sim}100$ minutes of ALMA light curves by time-averaging over ${\sim}1$ minute intervals, resulting in ${\sim}100$ data points for each Stokes component. Following the procedures outlined in \cite{wielgus2022orbital} we subtract a constant LP component with magnitude and angle of \mbox{$\bgmag = 0.16~\jansky$}, \mbox{$\bgangle = -37^\circ$} respectively, to account for the background accretion disk; We de-rotate the electric vector polarization angle (EVPA) by $32.2^\circ$ to account for the estimated Faraday rotation~\cite{wielgus2022orbital}. We model the data as homoscedatic within a short and stable observation window with a polarimetric noise level estimated at \mbox{$\sigma_Q = \sigma_U = 0.01~\jansky$}~\cite{wielgus2022orbital}. While we do not fit the total intensity, we regularize the model to have a total intensity around $0.3~\jansky$ with a standard deviation of $0.15~\jansky$~\cite{wielgus2022orbital}. Following \cite{wielgus2022orbital} we set 9:20 UT as the initial time of the analysis and 3D reconstruction of the flare. Supplementary Fig.~3 shows an analysis of different initial times around 9:20 UT, which provides further motivation for the selection of this initial time.

\subsection*{Forward model}
\label{sec:forwad}
In this section, we formulate the forward model, which takes a canonical 3D emission around a black hole as input and synthesizes light curves as output. Figure \ref{fig:overview} provides a high-level overview of the forward model, divided into four key building blocks, which we describe in the sections below.

\subsubsection*{1. Orbit dynamics}
\label{subsec:orbit}
The key assumption for orbital tomography is that the 4D (space and time) emission, $\emission(t,\bfx)$, is in orbit around the black hole and can be modeled as a simple transformation of a canonical (or initial) 3D emission, $\emission_0(\bfx)$:
\begin{equation}
    \emission(t, \bfx) = \emission_0({\bf T}_t \bfx),
\end{equation}
The transformation ${\bf T}_t$ propagates the initial 3D structure in time and connects dynamic observations, such as light curves, to the canonical 3D structure. This in turn enables formulating an inverse problem of estimating $\emission_0(\bfx)$ from time-variable observations. While the assumption of a coordinate transformation does not hold in general, it is well suited for compact, bright structures over short time scales, during which complex dynamics are negligible. 

In our work, we consider a Keplerian orbit model with an angular velocity:
\begin{equation}
\Omega(r) =\frac{\sqrt{M}}{r^{3/2} + a\sqrt{M}},
\label{eq:angular_velocity}
\end{equation}
where $r$ is the distance from the black-hole center and $M$ is the black-hole mass. Note that for $a=0$, Eq.~\eqref{eq:angular_velocity} coincides with the Newtonian expression for angular velocity. A purely azimuthal orbit is suitable outside the innermost stable circular orbit (ISCO), where radial velocities play a smaller role. Thus, we formulate the coordinate transformation as a shearing operation: 
\begin{equation}
    {\bf T}_t = \shear_{\phi},
\end{equation}
where $\shear_{\phi}$ is a rotation matrix at an angle:
\begin{equation}
\phi \left(t, r\right) = (t -t_0)\Omega(r).
\end{equation}
The angular velocity dependence on $r$ (Eq.~\ref{eq:angular_velocity}) causes shearing due to the faster motion of inner radii.

\subsubsection*{2. Image formation}
\label{subsec:image_formation}
In this section, we describe how $\emission_0$ relates to light curve observations through an image-formation model. Each image pixel collects radiation along a geodesic curve: $\Gamma(\bhparams, \alpha, \beta)$ terminating at the image coordinates $(\alpha, \beta)$. The ray path $\Gamma$ is determined by a handful of black hole parameters: $\bhparams$.
Omitting the explicit dependency on image coordinates (for brevity), we model image pixels through the polarized radiative transfer~\cite{GRay,grtrans,ipole} of an optically thin disk (attenuation can be neglected for \sgra{} 230GHz observations~\cite{akiyama2022firstV}):
\begin{equation}
    \pixel(t) =     
    \begin{bmatrix}
    p_I(t) \\ p_Q(t) \\ p_U(t) \\ p_V(t)
    \end{bmatrix}
    = \! \! \! \! \! \int \limits_{\bfx\in\Gamma(\bhparams)} \! \! \! \! \! \! \!g(\bfx)^2  \emission(t{+}\tau_\bfx, \bfx) \rotation (\bfx) {\bf J}(\bfx) d\bfx.
    \label{eq:rt}
\end{equation}
Equation \eqref{eq:rt} describes how pixel values are computed through an integration along geodesic curves $\Gamma$ computed by solving a set of differential equations~\cite{kgeo2023} (Supplementary Material, Sec.~3). The integrand in comprised of five elements: $g$, $\emission$, $\rotation$ and ${\bf J}$. $\emission(t{+}\tau_\bfx, \bfx)$ is the {\em unknown} scalar emissivity that depends on microphysical properties (e.g. local electron density and temperature) and $\tau_\bfx$ is the time delay that accounts for photon travel time (often referred to as slow-light). We model the polarized synchrotron radiation as this scalar emissivity function multiplied by a Stokes-vector ${\bf J}$ proportional to~\cite{gelles2021polarized}
\begin{align}
    J_I &\propto g^{\spectralindx}\left(|\magfield|\sin\pitch \right)^{\spectralindx+1}  \\ 
    J_Q &\propto q_f J_I \\ 
    J_U &= 0
    \label{eq:J_factors}
\end{align}
In this work, we consider only linear polarization, thus, setting $J_V=0$. Moreover, the spectral index (reflecting the change in the local emission with frequency) is approximated as $\spectralindx \simeq1$~\cite{Narayan2021}. Note that the local emission frame is defined to align with Stokes-$Q$, therefore, $J_U\equiv0$. The scaling factor $q_f \in [0,1]$ is the (volumetric) fraction of linear polarization and $\pitch$ is the angle between the local magnetic field $\bf B$ and photon momentum $\bf k$, given by
\begin{equation}
\sin\pitch(\bfx) = \frac{{\bf k}(\bfx) \times \magfield(\bfx)}{\left|\momentum(\bfx)\right|\left| \magfield(\bfx) \right|}
\end{equation}
The two remaining quantities to define are ${\bf R}$ and $g$. The matrix ${\bf R}$ rotates the LP, $(J_Q, J_U)$, from the emission frame to the image coordinates through {\em parallel-transport}~\cite{himwich2020universal} (see Supplementary Material Sec.~3.4). The scalar field $g(\bfx)$ is a General Relativistic (GR) red-shift factor, which decreases the emission when the material is deep in the gravitational field or moving away from the observer. More generally $g(\bfx)$ depends on the local direction of motion, $\velocity$, relative to the photon momentum ${\bf k}$
\begin{equation}
g(\bfx) = \big\langle \velocity(\bfx),\momentum(\bfx) \! \big\rangle
\end{equation}
Note that $\velocity,\momentum$ are 4-vectors, more explicitly defined in the Supplementary Material (Sec.~3).

\subsubsection*{3. Light curves}
For a given 3D emission, ray-tracing Eq.~\eqref{eq:rt} enables computing a single pixel value over time. We compute light curves by numerically sampling a large field-of-view (FOV) and summing over image-plane coordinates:
\begin{equation}
    \lc(t) =     
    \begin{bmatrix}
    I_I(t) \\ I_Q(t) \\ I_U(t) \\ I_V(t)
    \end{bmatrix} =
    \sum\limits_{\alpha, \beta} \pixel(t, \alpha, \beta)
    \label{eq:lc}
\end{equation}

\subsubsection*{4. Neural representation}
We formulate a tomographic recovery relying on a {\em neural representation}~\cite{levis2022gravitationally,mildenhall2020nerf} of the unknown 3D volume: $\emission_0(\bfx)$. Thus, instead of a traditional voxel discretization, the volume is represented by the weights, $\netparams$, of a multilayer perceptron (MLP), that are adjusted to fit the observations. 

The implicit regularization of the multilayer perceptron (MLP) architecture enables tackling highly ill-posed inverse problems~\cite{levis2022gravitationally, zhong2021cryodrgn}. The MLP takes continuously-valued coordinates $\bfx$ as input, and outputs the corresponding scalar emission at that coordinate:
\begin{equation}
    \emission_0 \left(\bfx \right) = \mlp_{\netparams}(\gamma(\bfx)),
    \label{eq:MLP}
\end{equation}
where $\gamma(\bfx)$ is a positional encoding of the input coordinates. 

Studies have shown~\cite{tancik2020fourfeat} that encoding the coordinates, instead of directly taking them as inputs, can capture continuous fields better (converging in the width limit to a stationary interpolation kernel~\cite{jacot2018neural}). Thus, our work relies on a positional encoding that projects each coordinate onto a set of sinusoids with exponentially increasing frequencies:
\begin{equation}
    \gamma(\bfx) {=} \Big[ \sin(\bfx), \cos(\bfx), \ldots 
    ,\sin\!\big(2^{L-1} \bfx\big), \cos\!\big(2^{L-1} \bfx\big) \Big]^T
\end{equation}
The positional encoding controls the underlying interpolation kernel used by the MLP, where the parameter $L$ determines the bandwidth of the interpolation kernel~\cite{tancik2020fourfeat}. 

In our work, we use a small MLP with 4 fully connected layers, where each layer is 128 units wide and uses ReLU activations. We use a maximum positional encoding degree of $L=3$. The low degree of $L$ is suitable for volumetric emission fields, which are naturally smooth~\cite{levis2022gravitationally}. 

Once the neural network weights, $\netparams$, are adjusted to fit the data, the network can be sampled at any 3D point, $\bfx$, to produce the emission value at that point. This allows us to: a) sample the network at regular grid points to extract a 3D volume representation of the recovered emission; b) ray trace the recovered emission as it would be seen by a perspective (pin-hole) camera (used for the visualizations throughout the paper).

\subsection*{Solving the inverse problem}
\label{subsec:inverse}
In this section, we formulate an optimization approach that enables jointly estimating the 3D emission and inclination, which are the parameters of the forward model. Supplementary Fig.~1 shows a high-level illustration of the data-fitting procedure introduced in the following section.

\subsubsection*{Tomographic reconstruction}
To estimate the 3D emission from light curve observations, we formulate a minimization problem. We estimate $\netparams$, which parameterizes $\emission_0(\bfx)$, by minimizing a $\chi^2$ data fit for each Stokes component, evaluated for a fixed set of black hole parameters, $\bhparams$:
\begin{equation}
    \mathcal{\chi}^2(\netparams|\bhparams) = \mathcal{\chi}^2_I(\netparams|\bhparams) + \mathcal{\chi}^2_Q(\netparams|\bhparams) + \mathcal{\chi}^2_U(\netparams|\bhparams).
    \label{eq:chisq}
\end{equation}
Here, we restrict the discussion to the total intensity and LP components: $I, Q, U$. Each $\chi^2$ is calculated as a sum over discrete temporal data points
\begin{equation}
    \mathcal{\chi}^2_s(\netparams|\bhparams) = \frac{1}{N_{\rm obs}}\sum \limits_i \left( \frac{\measurements_s(t_i) - I_s(t_i, \netparams|\bhparams)}{\sigma_s} \right)^2,
    \label{eq:chisq_stokes}
\end{equation}
where $N_{\rm obs}$ is the total number of data points and the subscript $s=\left\{ I, Q, U, V \right\}$ represents the stokes components, $\measurements_s$, $I_s$ and $\sigma_s$ are the polarimetric observations, model, and noise standard deviation, respectively. Note that $I_s(t_i, \netparams|\bhparams)$ is simply the light curve given by Eq.~\eqref{eq:lc}, sampled at discrete time $t_i$, where $\netparams|\bhparams$ highlight its dependency/conditioning on the network/black hole parameters.

Equation \eqref{eq:chisq} depends on {\em unknown} black hole parameters; nevertheless, the mass of \sgra{} can be constrained through stellar dynamics~\cite{ghez2008measuring,abuter2022mass}; \mbox{$\mass\simeq 4 {\times} 10^6 \solmass$} where $\solmass$ denotes solar masses. Furthermore, since the data fit is insensitive to black hole spin, the only estimated parameter is the inclination angle. To estimate the inclination, we numerically bin $\inc \in [0,\pi/2]$ and recover the 3D emission by minimizing Eq.~\eqref{eq:chisq}:
\begin{equation}
\netparams^\star (\inc) = \arg \min_\netparams \chi^2\left(\netparams|\inc \right).
\label{eq:minimization}
\end{equation}
By interpreting Eq.~\eqref{eq:minimization} as a function of $\inc$ we approximate the marginal log-likelihood as
\begin{equation}
\mathcal{L} (\inc|\measurements) \propto \chi^2(\inc|\netparams^\star).
\label{eq:theta_likelihood}
\end{equation}
Equation \eqref{eq:theta_likelihood} is a zero-order expansion about the maximum likelihood estimator: $\netparams^\star$.

\subsubsection*{Model selection using Validation-$\chi^2$}
While Eq.~\eqref{eq:theta_likelihood} tells us how well each model (inclination) fits the data, it is susceptible to over-fitting. To mitigate over-fitting, we define a more robust metric called {\em validation-$\chi^2$} (see Supplementary Fig.~5). The inclination angle is then estimated through the following procedure: 
\begin{enumerate}[topsep=0pt]
  \item During optimization, ray positions are fixed to the center of each image pixel. In our recoveries, we use an evenly sampled $64\times64$ grid for a FOV of $200~\uas$.
  \item We compute $\chi^2$ for perturbed pixel positions (off-center) within a small pixel area. In our recoveries, we used a pixel area of $3.125 \times 3.125~\uas^2$.
  \item We average $\chi^2$ of 10 randomly sampled (uniform) ray positions to compute validation-$\chi^2$ curves.
  \item $\inc^\star$ is estimated as the global minimum of the validation $\chi^2$.
\end{enumerate}
Through simulations, we highlight how this procedure is a more robust selection criterion for models that are not overfitting the fixed ray positions (Supplementary Fig.~5). 

\subsubsection*{Optimization procedure}
The neural network was implemented in {\tt JAX}~\cite{jax2018github}. Both the synthetic experiments (Supplementary Material, Sec.~2) and the ALMA recovery were optimized using an ADAM optimizer~\cite{kingma2014adam} with a polynomial learning rate transitioning from $1{\rm e}^{-4} \rightarrow 1{\rm e}^{-6}$ over $50{\rm K}$ iterations. Run times were ${\sim} 1$ hour on two NVIDIA Titan RTX GPUs. Network weights were randomly initialized (Gaussian distributed) with several initial seeds.

\section*{Data Availability}
This paper makes use of the ALMA data set ADS/JAO.ALMA\#2016.1.01404.V, available through the ALMA data portal. Fully calibrated data and other materials are available from the corresponding author upon reasonable request.

\section*{Code Availability}
The  software  packages  used  analyze  the  data  are  available  at  the  following sites: {\tt kgeo} (\url{https://github.com/achael/kgeo}), {\tt bhnerf} (\url{https://github.com/aviadlevis/bhnerf}).

\section*{Acknowledgements}
We would like to thank Vikram Ravi and Bart Ripperda for fruitful discussions. This work was supported by NSF awards 1935980, 2048237, and the Carver Mead New Adventures Fund. A.~A.~C. is supported by the Princeton Gravity Initiative. M.~W.'s research is supported by the ERC advanced grant ``M2FINDERS - Mapping Magnetic Fields with INterferometry Down to Event hoRizon Scales'' (Grant No. 101018682)

\section*{Author Contributions}
A.~L. and K.~L.~B. identified and formulated the tomography problem and conceived and designed the experiments. A.~L., K.~L.~B., and P.~P.~S. developed the neural representation-based tomography approach. A.~A.~C. and A.~L. developed the polarimetric ray tracing. A.~A.~C formulated and implemented the geodesic computations. M.~W. worked on data calibration. A.~L. performed the tomographic reconstructions. All authors contributed to the analysis, interpretation, discussions, and writing of the manuscript.

\section*{Competing interests}
The authors declare no competing interests.

\bibliographystyle{plain}

\begin{thebibliography}{10}

\bibitem{genzel2003near}
R.~Genzel, R.~Sch{\"o}del, T.~Ott, A.~Eckart, T.~Alexander, F.~Lacombe, D.~Rouan, and B.~Aschenbach.
\newblock Near-infrared flares from accreting gas around the supermassive black hole at the Galactic Centre.
\newblock {\em Nature}, 425(6961):934-937, 2003.

\bibitem{neilsen2013chandra}
J.~Neilsen, M.~A.~Nowak, C.~Gammie, J.~Dexter, S~Markoff, D.~Haggard, S.~Nayakshin, Q.~D.~Wang, N.~Grosso, D.~Porquet, et~al.
\newblock A Chandra/HETGS Census of X-ray Variability From Sgr A$^{*}$ during 2012.
\newblock {\em \apj}, 774(1):42, 2013.

\bibitem{fazio2018multiwavelength}
G.~G.~Fazio, J.~L.~Hora, G.~Witzel, S.~P.~Willner, M.~L.~N.~Ashby, F.~Baganoff, E.~Becklin, S.~Carey, D.~Haggard, C.~Gammie, et~al.
\newblock Multiwavelength Light Curves of Two Remarkable Sagittarius A$^{*}$ Flares.
\newblock {\em \apj}, 864(1):58, 2018.

\bibitem{Wielgus2022lc}
M.~{Wielgus}, N.~{Marchili}, I.~{Mart{\'\i}-Vidal}, G.~K.~{Keating}, V.~{Ramakrishnan}, P.~{Tiede}, E.~{Fomalont}, S.~{Issaoun}, J.~{Neilsen}, M.~A.~{Nowak}, et~al.
\newblock {Millimeter Light Curves of Sagittarius A$^{*}$ Observed during the 2017 Event Horizon Telescope Campaign}.
\newblock {\em \apjl}, 930(2):L19, 2022.

\bibitem{Marrone_2008}
D.~P.~{Marrone}, F.~K.~{Baganoff}, M.~R.~{Morris}, J.~M.~{Moran}, A.~M.~{Ghez}, S.~D.~{Hornstein}, C.~D.~{Dowell}, D.~J.~{Mu{\~n}oz}, M.~W.~{Bautz}, G.~R.~{Ricker}, W.~N.~{Brandt}, G.~P.~{Garmire}, J.~R.~{Lu}, K.~{Matthews}, J.~H.~{Zhao}, R.~{Rao}, and G.~C.~{Bower}.
\newblock {An X-Ray, Infrared, and Submillimeter Flare of Sagittarius A$^{*}$}.
\newblock {\em \apj}, 682(1):373--383, 2008.

\bibitem{Haggardetal2019}
D.~{Haggard}, M.~{Nynka}, B.~{Mon}, N.~{de la Cruz Hernandez}, M.~{Nowak}, C.~{Heinke}, J.~{Neilsen}, J.~{Dexter}, P.~Chris {Fragile}, F.~{Baganoff}, G.~C.~{Bower}, L.~R.~{Corrales}, F.~{Coti Zelati}, N.~{Degenaar}, S.~{Markoff}, M.~R.~{Morris}, G.~{Ponti}, N.~{Rea}, J.~{Wilms}, and F.~{Yusef-Zadeh}.
\newblock {Chandra Spectral and Timing Analysis of Sgr A$^{*}$'s Brightest X-Ray Flares}.
\newblock {\em \apj}, 886(2):96, 2019.

\bibitem{dexter2020sgr}
J.~{Dexter}, A.~{Tchekhovskoy}, A.~{Jim{\'e}nez-Rosales}, S.~M. {Ressler}, M.~{Baub{\"o}ck}, Y.~{Dallilar}, P.~T. {de Zeeuw}, F.~{Eisenhauer}, S.~{von Fellenberg}, F.~{Gao}, R.~{Genzel}, S.~{Gillessen}, M.~{Habibi}, T.~{Ott}, J.~{Stadler}, O.~{Straub}, and F.~{Widmann}.
\newblock {Sgr A$^{*}$ near-infrared flares from reconnection events in a magnetically arrested disc}.
\newblock {\em \mnras}, 497(4):4999--5007, 2020.

\bibitem{Witzel_2021}
G.~{Witzel}, G.~{Martinez}, S.~P.~{Willner}, E.~E.~{Becklin}, H.~{Boyce}, T.~{Do}, A.~{Eckart}, G.~G.~{Fazio}, A.~{Ghez}, M.~A.~{Gurwell}, D.~{Haggard}, R.~{Herrero-Illana}, J.~L.~{Hora}, Z.~{Li}, J.~{Liu}, N.~{Marchili}, M.~R.~{Morris}, H.~A.~{Smith}, M.~{Subroweit}, and J.~A.~{Zensus}.
\newblock {Rapid Variability of Sgr A$^{*}$ across the Electromagnetic Spectrum}.
\newblock {\em \apj}, 917(2):73, 2021.

\bibitem{broderick2005imaging}
A.~E.~{Broderick} and A.~{Loeb}.
\newblock {Imaging bright-spots in the accretion flow near the black hole horizon of Sgr A$^{*}$}.
\newblock {\em \mnras}, 363(2):353--362, 2005.

\bibitem{ripperda2022black}
B.~Ripperda, M.~Liska, K.~Chatterjee, G.~Musoke, A.~A.~Philippov, S.~B.~Markoff, A.~Tchekhovskoy, and Z.~Younsi.
\newblock Black hole flares: ejection of accreted magnetic flux through 3{D} plasmoid-mediated reconnection.
\newblock {\em \apjl}, 924(2):L32, 2022.

\bibitem{gravity18}
{GRAVITY Collaboration}.
\newblock {Detection of orbital motions near the last stable circular orbit of the massive black hole SgrA$^{*}$}.
\newblock {\em \aap}, 618:L10, 2018.

\bibitem{gravity20}
{GRAVITY Collaboration}.
\newblock {Modeling the orbital motion of Sgr A$^{*}$'s near-infrared flares}.
\newblock {\em \aap}, 635:A143, 2020.

\bibitem{wielgus2022orbital}
M.~{Wielgus}, M.~{Moscibrodzka}, J.~{Vos}, Z.~{Gelles}, I.~{Mart{\'\i}-Vidal}, J.~{Farah}, N.~{Marchili}, C.~{Goddi}, and H.~{Messias}.
\newblock {Orbital motion near Sagittarius A$^{*}$ . Constraints from polarimetric ALMA observations}.
\newblock {\em \aap}, 665:L6, 2022.

\bibitem{akiyama2022firstI}
{Event Horizon Telescope Collaboration}.
\newblock {First Sagittarius A$^{*}$ Event Horizon Telescope Results. I. The Shadow of the Supermassive Black Hole in the Center of the Milky Way}.
\newblock {\em \apjl}, 930(2):L12, 2022.

\bibitem{akiyama2022firstVI}
{Event Horizon Telescope Collaboration}.
\newblock {First Sagittarius A$^{*}$ Event Horizon Telescope Results. VI. Testing the Black Hole Metric}.
\newblock {\em \apjl}, 930(2):L17, 2022.

\bibitem{akiyama2022firstIII}
{Event Horizon Telescope Collaboration}.
\newblock {First Sagittarius A$^{*}$ Event Horizon Telescope Results. III. Imaging of the Galactic Center Supermassive Black Hole}.
\newblock {\em \apjl}, 930(2):L14, 2022.

\bibitem{akiyama2022firstII}
{Event Horizon Telescope Collaboration}.
\newblock {First Sagittarius A$^{*}$ Event Horizon Telescope Results. II. EHT and Multiwavelength Observations, Data Processing, and Calibration}.
\newblock {\em \apjl}, 930(2):L13, 2022.

\bibitem{Tiede_2020}
P.~{Tiede}, H.~Y.~{Pu}, A.~E.~{Broderick}, R.~{Gold}, M.~{Karami}, and J.~A.~{Preciado-L{\'o}pez}.
\newblock {Spacetime Tomography Using the Event Horizon Telescope}.
\newblock {\em \apj}, 892(2):132, 2020.

\bibitem{levis2021inference}
A.~Levis, D.~Lee, J.~A.~Tropp, C.~F.~ Gammie, and K.~L.~Bouman.
\newblock Inference of Black Hole Fluid-Dynamics from Sparse Interferometric Measurements.
\newblock In {\em Proc. IEEE/CVF Intl. Conf. Comput. Vision (ICCV)}, 2340--2349, 2021.

\bibitem{levis2022gravitationally}
A.~Levis, P.~P.~Srinivasan, A.~A.~Chael, R.~Ng, and K.~L.~Bouman.
\newblock Gravitationally Lensed Black Hole Emission Tomography.
\newblock In {\em Proc. IEEE/CVF Conf. Comput. Vis. Pattern Recognit. (CVPR)}, 19841--19850, 2022.

\bibitem{mildenhall2020nerf}
B.~Mildenhall, P.~P.~Srinivasan, M.~Tancik, J.~T.~Barron, R.~Ramamoorthi, and R.~Ng.
\newblock NeRF: Representing Scenes as Neural Radiance Fields for View Synthesis.
\newblock {\em Eur. Conf. Comput. Vis. (ECCV)}, 2020.

\bibitem{tancik2020fourfeat}
M.~Tancik, P.~P.~Srinivasan, B.~Mildenhall, S.~F.~Keil, N.~Raghavan, U.~Singhal, R.~Ramamoorthi, J.~T.~Barron, and R.~Ng.
\newblock Fourier Features Let Networks Learn High Frequency Functions in Low Dimensional Domains.
\newblock {\em Adv. Neural Inf. Process. Syst. (NeurIPS)}, 33:7537-7547, 2020.

\bibitem{Gravity2023}
{Gravity Collaboration}.
\newblock {Polarimetry and astrometry of NIR flares as event horizon scale dynamical probes for the mass of Sgr A$^{*}$}.
\newblock {\em \aap}, 677:L10, 2023.

\bibitem{gelles2021polarized}
Z.~Gelles, E.~Himwich, M.~D.~Johnson, and D.~C.~M.~Palumbo.
\newblock Polarized image of equatorial emission in the Kerr geometry.
\newblock {\em \prd}, 104(4):044060, 2021.

\bibitem{vos2022polarimetric}
J.~{Vos}, M.~A. {Mo{\'s}cibrodzka}, and M.~{Wielgus}.
\newblock {Polarimetric signatures of hot spots in black hole accretion flows}.
\newblock {\em \aap}, 668:A185, 2022.

\bibitem{yfantis2023}
A.~I.~{Yfantis}, M.~A.~{Mościbrodzka}, M.~{Wielgus}, J.~T.~{Vos}, A.~{Jimenez-Rosales}.
\newblock {Fitting Sagittarius A$^{*}$ light curves with a hot spot model: Bayesian modeling of QU loops in millimeter band}.
\newblock {\em arXiv preprint}, 2310.07762, 2023.

\bibitem{akiyama2022firstV}
{Event Horizon Telescope Collaboration}.
\newblock {First Sagittarius A$^{*}$ Event Horizon Telescope Results. V. Testing Astrophysical Models of the Galactic Center Black Hole}.
\newblock {\em \apjl}, 930(2):L16, 2022.

\bibitem{ghez2008measuring}
A.~M.~Ghez, S.~Salim, N.~N.~Weinberg, J.~R.~Lu, T.~Do, J.~K.~Dunn, K.~Matthews, M.~R.~Morris, S.~Yelda, E.~E.~Becklin, et~al.
\newblock Measuring Distance and Properties of the Milky Way's Central Supermassive Black Hole with Stellar Orbits.
\newblock {\em \apj}, 689(2):1044, 2008.

\bibitem{akiyama2019firstIV}
{Event Horizon Telescope Collaboration}.
\newblock {First M87 Event Horizon Telescope Results. IV. Imaging the Central Supermassive Black Hole}.
\newblock {\em \apjl}, 875(1):L4, 2019.

\bibitem{Issaoun2019}
S.~{Issaoun}, M.~D.~{Johnson}, L.~{Blackburn}, C.~D.~~{Brinkerink}, M.~{Mościbrodzka}, A.~{Chael}, C.~{Goddi}, I.~{Martí-Vidal}, J.~ {Wagner}, S.~S.~{Doeleman}, H~{Falcke}.
\newblock {The size, shape, and scattering of Sagittarius A$^{*}$ at 86 GHz: first VLBI with ALMA}.
\newblock {\em \apj}, 871(1):30, 2019.

\bibitem{cho2022}
{I.~Cho, G.~Y.~Zhao, T.~Kawashima, M.~Kino, K.~Akiyama,  M.~D.~Johnson et al.}.
\newblock {The Intrinsic Structure of Sagittarius A$^{*}$ at 1.3 cm and 7 mm}.
\newblock {\em \apjl}, 926(2):108, 2022.

\bibitem{GRay}
CK.~{Chan}, D.~{Psaltis}, and F.~{{\"O}zel}.
\newblock {GRay: A Massively Parallel GPU-based Code for Ray Tracing in Relativistic Spacetimes}.
\newblock {\em \apj}, 777(1):13, 2013.

\bibitem{grtrans}
J.~{Dexter}.
\newblock {A public code for general relativistic, polarised radiative transfer around spinning black holes}.
\newblock {\em \mnras}, 462(1):115--136, 2016.

\bibitem{ipole}
M.~{Mo{\'s}cibrodzka} and C.~F. {Gammie}.
\newblock {{IPOLE} - semi-analytic scheme for relativistic polarized radiative transport}.
\newblock {\em \mnras}, 2018.

\bibitem{kgeo2023}
A~Chael.
\newblock kgeo (version 1.0.0).
\newblock \url{https://github.com/achael/kgeo}, 2023.
\newblock Online; accessed 01-June-2023.

\bibitem{Narayan2021}
R.~{Narayan}, D.~C.~M.~ {Palumbo}, M.~D.~{Johnson}, Z.~{Gelles}, E.~{Himwich}, D.~O.~{Chang}, A.~{Ricarte}, J.~{Dexter}, C.~F.~{Gammie}, A.~A. {Chael}, and {Event Horizon Telescope Collaboration}.
\newblock {The Polarized Image of a Synchrotron-emitting Ring of Gas Orbiting a Black Hole}.
\newblock {\em \apjl}, 912(1):35, 2021.

\bibitem{himwich2020universal}
E.~Himwich, M.~D.~Johnson, A.~Lupsasca, and A.~Strominger.
\newblock Universal polarimetric signatures of the black hole photon ring.
\newblock {\em \prd}, 101(8):084020, 2020.

\bibitem{zhong2021cryodrgn}
E.~D.~Zhong, T.~Bepler, B.~Berger, and J.~H.~Davis.
\newblock CryoDRGN: reconstruction of heterogeneous cryo-EM structures using neural networks.
\newblock {\em Nat. Methods}, 18(2):176--185, 2021.

\bibitem{jacot2018neural}
A.~Jacot, F.~Gabriel, and C.~Hongler.
\newblock Neural Tangent Kernel: Convergence and Generalization in Neural Networks.
\newblock {\em Adv. Neural Inf. Process. Syst. (NeurIPS)}, 31, 2018.

\bibitem{abuter2022mass}
R~Abuter, N~Aimar, A~Amorim, J~Ball, M~Baub{\"o}ck, JP~Berger, H~Bonnet, G~Bourdarot, W~Brandner, V~Cardoso, et~al.
\newblock Mass distribution in the galactic center based on interferometric astrometry of multiple stellar orbits.
\newblock {\em \aap}, 657:L12, 2022.

\bibitem{jax2018github}
J.~Bradbury, R.~Frostig, P.~Hawkins, M.~J.~Johnson, C.~Leary, D.~Maclaurin, G.~Necula, A.~Paszke, J.~Vander{P}las, S.~Wanderman-{M}ilne, and Q.~Zhang.
\newblock {JAX}: composable transformations of {P}ython+{N}um{P}y programs,
  2018.
\bibitem{kingma2014adam}
D.~P.~Kingma and J.~Ba.
\newblock Adam: A Method for Stochastic Optimization.
\newblock In {\em International Conference for Learning Representations (ICLR)}, 2015.

\end{thebibliography}

\begin{thebibliography}{10}

\bibitem{wielgus2022orbital_supp}
M.~{Wielgus}, M.~{Moscibrodzka}, J.~{Vos}, Z.~{Gelles}, I.~{Mart{\'\i}-Vidal}, J.~{Farah}, N.~{Marchili}, C.~{Goddi}, and H.~{Messias}.
\newblock {Orbital motion near Sagittarius A$^{*}$ . Constraints from polarimetric ALMA observations}.
\newblock {\em \aap}, 665:L6, 2022.

\bibitem{akiyama2022firstIII_supp}
{Event Horizon Telescope Collaboration}.
\newblock {First Sagittarius A$^{*}$ Event Horizon Telescope Results. III. Imaging of the Galactic Center Supermassive Black Hole}.
\newblock {\em \apjl}, 930(2):L14, 2022.

\bibitem{levis2021inference_supp}
A.~Levis, D.~Lee, J.~A.~Tropp, C.~F.~ Gammie, and K.~L.~Bouman.
\newblock Inference of Black Hole Fluid-Dynamics from Sparse Interferometric Measurements.
\newblock In {\em Proc. IEEE/CVF Intl. Conf. Comput. Vision (ICCV)}, 2340--2349, 2021.

\bibitem{lee2021disks_supp}
D.~Lee and C.~F.~Gammie.
\newblock Disks as Inhomogeneous, Anisotropic Gaussian Random Fields.
\newblock {\em \apj}, 906(1):39, 2021.

\bibitem{akiyama2022firstI_supp}
{Event Horizon Telescope Collaboration}.
\newblock {First Sagittarius A$^{*}$ Event Horizon Telescope Results. I. The Shadow of the Supermassive Black Hole in the Center of the Milky Way}.
\newblock {\em \apjl}, 930(2):L12, 2022.

\bibitem{gralla2020null_supp}
S.~E.~Gralla and A.~Lupsasca.
\newblock Null geodesics of the Kerr exterior.
\newblock {\em \prd}, 101(4):044032, 2020.

\bibitem{kgeo2023_supp}
A~Chael.
\newblock kgeo (version 1.0.0).
\newblock \url{https://github.com/achael/kgeo}, 2023.
\newblock Online; accessed 01-June-2023.

\bibitem{himwich2020universal_supp}
E.~Himwich, M.~D.~Johnson, A.~Lupsasca, and A.~Strominger.
\newblock Universal polarimetric signatures of the black hole photon ring.
\newblock {\em \prd}, 101(8):084020, 2020.

\bibitem{grtrans_supp}
J.~{Dexter}.
\newblock {A public code for general relativistic, polarised radiative transfer around spinning black holes}.
\newblock {\em \mnras}, 462(1):115--136, 2016.

\bibitem{gelles2021polarized_supp}
Z.~Gelles, E.~Himwich, M.~D.~Johnson, and D.~C.~M.~Palumbo.
\newblock Polarized image of equatorial emission in the Kerr geometry.
\newblock {\em \prd}, 104(4):044060, 2021.

\end{thebibliography}

\clearpage

\setcounter{page}{1}
\setcounter{equation}{0}
\setcounter{figure}{0}

\title{Orbital Polarimetric Tomography of a Flare Near the Sagittarius A$^*$ Supermassive Black Hole: Supplementary Material}

\twocolumn[
  \begin{@twocolumnfalse}
    \maketitle
  \end{@twocolumnfalse}
]

\captionsetup[figure]{name={Supplementary Fig.},labelsep=period}
 
In this work, we formulate a novel approach for estimating an evolving 3D structure around a black hole from light curve observations. To solve a tomography from a single viewpoint, we rely on both radiative emission and gravitational physics around the black hole. Integrating these physical models with a neural representation of the unknown 3D structure enables solving a highly ill-posed inverse problem, which we term {\em orbital polarimetric tomography}. By scanning over a range of possible inclination angles, we jointly estimate the inclination angle and 3D structure that best fit the polarimetric light curve data. Supplementary Fig.~\ref{fig:datafit_illustration} shows an overview of the data-fitting procedure. 

In the supplementary material we a) detail our evaluation of the orbital polarimetric tomography on synthetic simulations and b) add additional details and context for the general relativistic ray tracing. The organization is as follows: Sec \ref{sec:synthetic_gen} describes the data generation, Sec.~\ref{sec:synthetic_rec} analyses the recovery results, and Sec.~\ref{sec:GR} covers additional details for general relativistic ray tracing, not included in the main text.

\begin{figure*}[t]
	\centering \includegraphics[width=\linewidth]{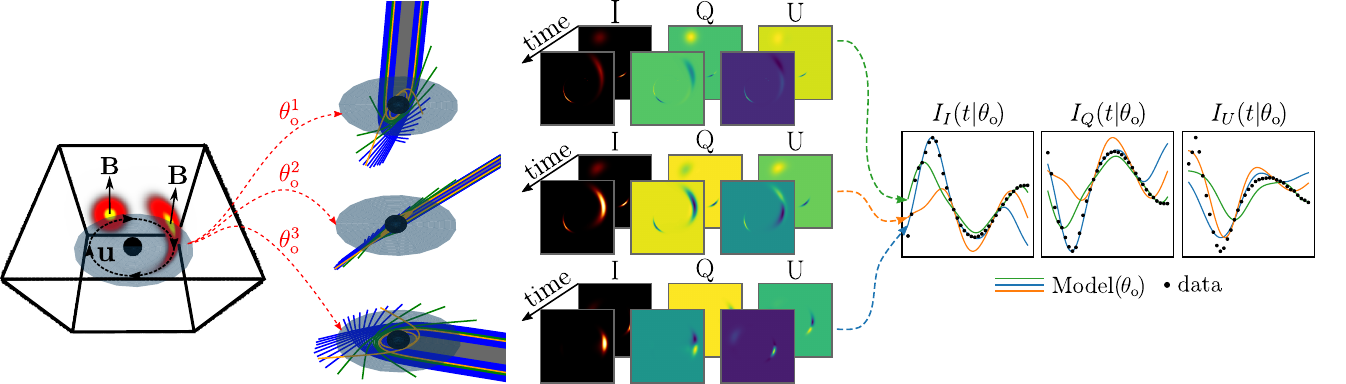}
	\caption{Illustration of the simultaneous estimation of 3D emission and inclination angle from polarimetric lightcurve data. The light curves ($I_I, I_Q, I_U$) depend on the unknown inclination angle ($\inc$) which gives rise to a different image-plane and subsequently model fit. From left to right: 1. An emission bright spot with azimuthal velocity $\velocity$ orbits a black hole on the equatorial plane. The vertical magnetic field $\magfield$ induces polarized synchrotron radiation. 2. Geodesic curves are computed for every pixel at each inclination angle. 3. Integrating emission along geodesics to generate a Stokes image plane over time. 4. Summing over image pixels gives the model prediction per inclination angle.
 }
	\label{fig:datafit_illustration}
\end{figure*}

\section{Synthetic data generation}
\label{sec:synthetic_gen}
To quantitatively assess our approach and the ability to recover the 3D structures and constrain an unknown inclination angle we generate synthetic data mimicking ALMA observations. Synthetic simulations allow for careful control of the sources of variability to evaluate the overall impact on the recovery performance. Each dataset consists of a flare on top of a background accretion disk as described in the following sections.

\subsection{Emission flares}
\begin{figure*}[t!]
	\centering \includegraphics[width=0.95\linewidth]{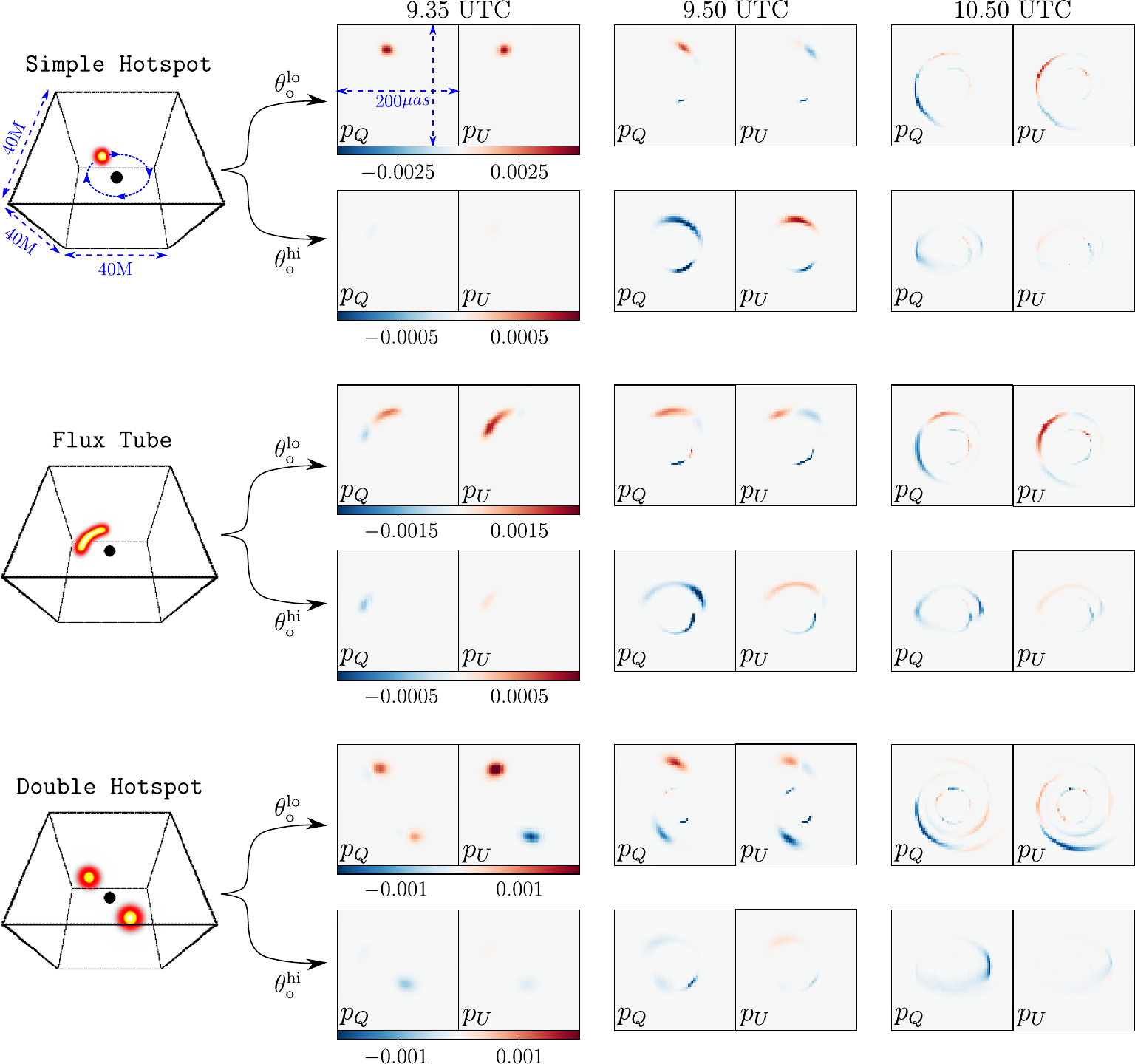}
	\caption{Synthetic simulations of three emission structures: {\tt Simple Hotspot}, {\tt Flux Tube}, {\tt Double Source}. Observations are generated at two inclination angles: $\inclo=12^\circ$, $\inchi=64^\circ$. This figure illustrates the gravitationally lensed LP images: $p_Q, p_U$ at three times within a period mimicking the {\em radio-loops} observed by ALMA: 9.35 -- 11 UTC (April 11, 2017). Images are computed at ${\rm FOV}=40M$ ($\sim 200 \uas$ for \sgra{}) discretized into $64{\times}64$ pixels. To reduce aliasing effects, each pixel is computed as an average of 10 randomly sampled sub-pixel rays. All models share the same dimensions and orbit highlighted in blue for the {\tt Simple Hotspot}. Note the secondary images formed by gravitational lensing, evident in the {\tt Simple Hotspot} and {\tt Flux Tube} models (the bottom half of the image plane at 9.50 UTC)}
	\label{fig:3d_cases}
\end{figure*}
\begin{figure}[t]
	\centering \includegraphics[width=0.9\linewidth]{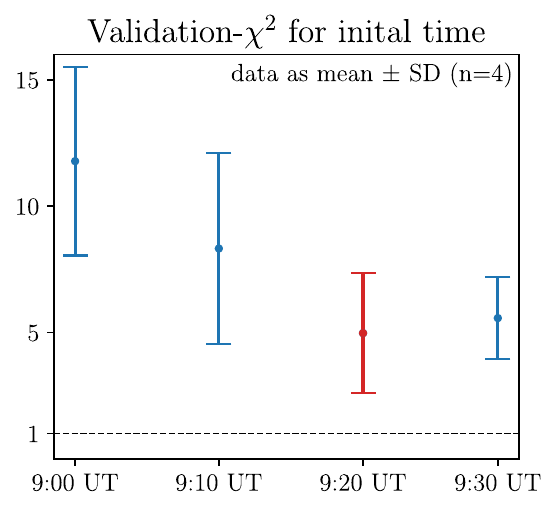}
        \caption{Validation-$\chi^2$ for different initial times around the 9:20 UT. The best data-fit (lowest validation-$\chi^2$) is highlighted in red. The error bars for each initial time are computed by recovering the 3D structure with four random initializations.}
	\label{fig:chi2_init_t}
\end{figure}

We generate data consisting of three distributions of emission: {\tt Simple Hotspot},  {\tt Flux Tube}, {\tt Double Source} (Supplementary Fig.~\ref{fig:3d_cases}). Each volume spans a $40 \times 40 \times 40~[{\rm M}^3]$ cube with emission occupying an equatorial disk of $\sim 4-6M$ thickness. For these simulations, we assume a non-spinning black hole with a constant vertical magnetic field $\magfield (\bfx)=\hat{\bf z}B_{z}$. We model flares with high fractional polarization resulting in average fluxes matching the real ALMA data~\cite{wielgus2022orbital_supp} 
\begin{align}
    \bar{I}_I &\simeq 0.3-0.4 ~\jansky, \label{eq:flux_req_I} \\
    \bar{P} &= \sqrt{I_Q^2 + I_U^2} \simeq 0.1-0.15 ~\jansky.
    \label{eq:flux_req_P}
\end{align}
Image frames and subsequent light curves are computed by discretizing an FOV of $40M$ (${\sim} 200 \uas$ for \sgra{}) into $64{\times}64$ pixels. To reduce aliasing, each pixel value is computed by ray-tracing and averaging 10 sub-pixel rays, sampled uniformly within the pixel fov. For each 3D structure, observations are generated at two inclination angles: $\inclo=12^\circ$ and $\inchi=64^\circ$ resulting in a dataset of six polarimetric light curve observations. Supplementary Fig.~\ref{fig:3d_cases} shows the 3D geometry and a sample of ray traced frames. 

Based on our analysis of the validation-$\chi^2$ for different initial times around 9:20 UT, we set the initial time of the flare to 9:20 UT, where the $\chi^2$ of the ALMA data-fitting is lowest (Supplementary Fig.~\ref{fig:chi2_init_t}). For each simulated emission structure, we synthesize $\sim 100$ minutes of observations to mimic the time-averaged ALMA observations. We use a sampling cadence of $\sim1$ data point per minute with appropriate scan gaps taken from the observational data. 

To produce fluxes that are comparable to ALMA observed fluxes, we adjust the magnitude of the 3D emission structure and the volumetric LP fraction to satisfy Eqs.~\eqref{eq:flux_req_I}--\eqref{eq:flux_req_P}. We use a $q_f=0.85$ to generate observations at $\inclo$ and $q_f=0.5$ for $\inchi$; A lower $q_f$ is used at a high inclination to account for the stronger Doppler boosting.

\subsection{Accretion disk}
\begin{figure*}[t]
	\centering \includegraphics[width=0.95\linewidth]{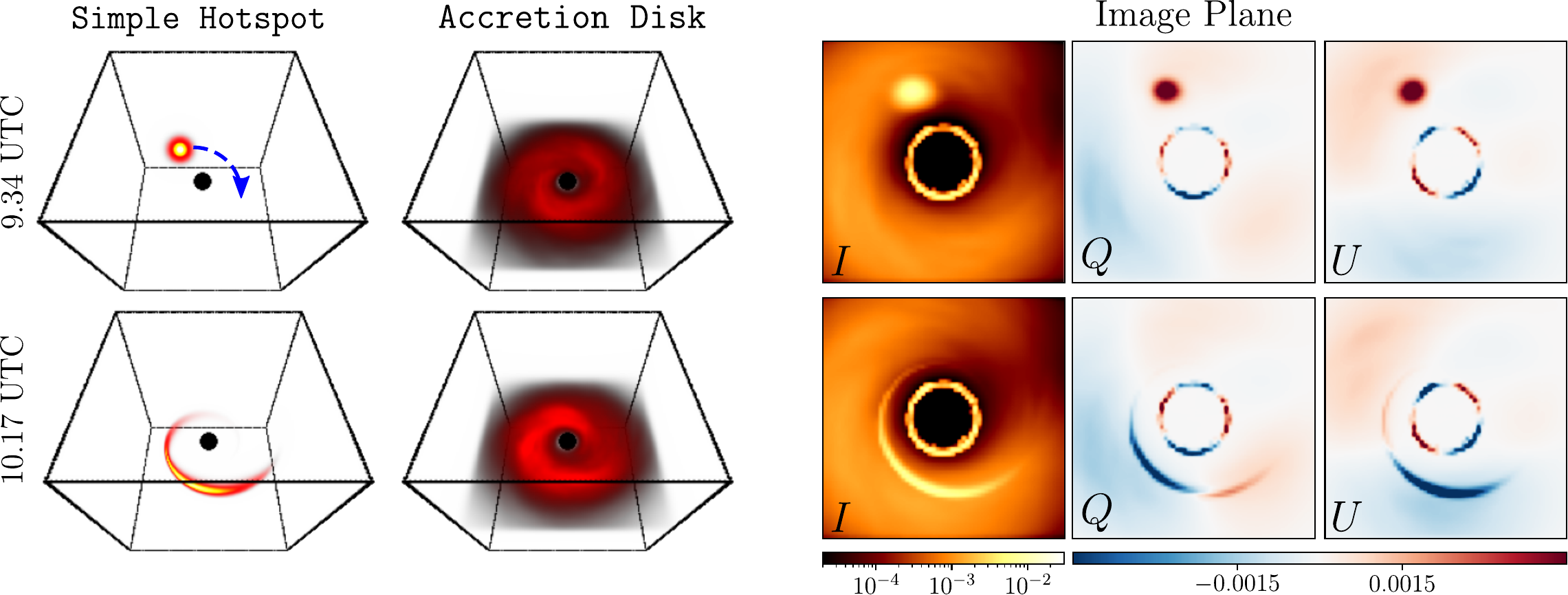}
	\caption{An illustration of the {\tt Simple Hotspot} and {\tt Accretion Disk} used for background noise. [left panels] Evolving 3D emission at three different times (rows). [right panels] Corresponding image frames ray-traced for a non-spinning black hole at a low inclination angle ($\inclo$). The bright ring is caused by gravitational lensing of the accretion disk with a diameter of $\sim50~\uas$ consistent with the images of \sgra{}~\cite{akiyama2022firstI_supp}.}
\label{fig:accretion_illustration}
\end{figure*}
While ALMA observations have low instrumental noise ($\sim2$ orders of magnitude less than the signal), a significant source of systematic noise is the unaccounted-for variability of the background accretion disk. To assess the impact the background signal has on reconstructions we generate synthetic data of a dynamic accretion disk. We model the accretion disk as a stochastic flow designed to mimic the statistics around \sgra{}~\cite{akiyama2022firstIII_supp, levis2021inference_supp}. The 2D spiral flow is generated as a Gaussian Random Field (GRF)~\cite{lee2021disks_supp} which is inflated to 3D through convolution with a 1D vertical Gaussian kernel with a full-width half max (FWHM) of $2.35{\rm M}$. Subsequently, the 3D flow is gravitationally lensed (with the same ray-tracing procedure described for the foreground emission flares) to generate a polarized image plane. Supplementary Fig.~\ref{fig:accretion_illustration} illustrates the 3D evolution of a randomly sampled {\tt Accretion Disk} alongside the {\tt Simple Hotspot}. Furthermore, the image plane containing both components is shown at two different evolution times. 

\section{Synthetic data recovery}
\label{sec:synthetic_rec}
In this section, we study the recovery performance through experiments designed to highlight different aspects of the tomography approach. In all recoveries, the 3D structure is visualized by sampling the network on a 3D grid at a resolution of $\sim 0.15{\rm M}$. The visualized volumes are in-fact the estimated intrinsic 3D emission (without spacetime curvature).

\subsection{Estimating inclination}
\label{subsec:est_inc}
\begin{figure}[t]
	\centering \includegraphics[width=\linewidth]{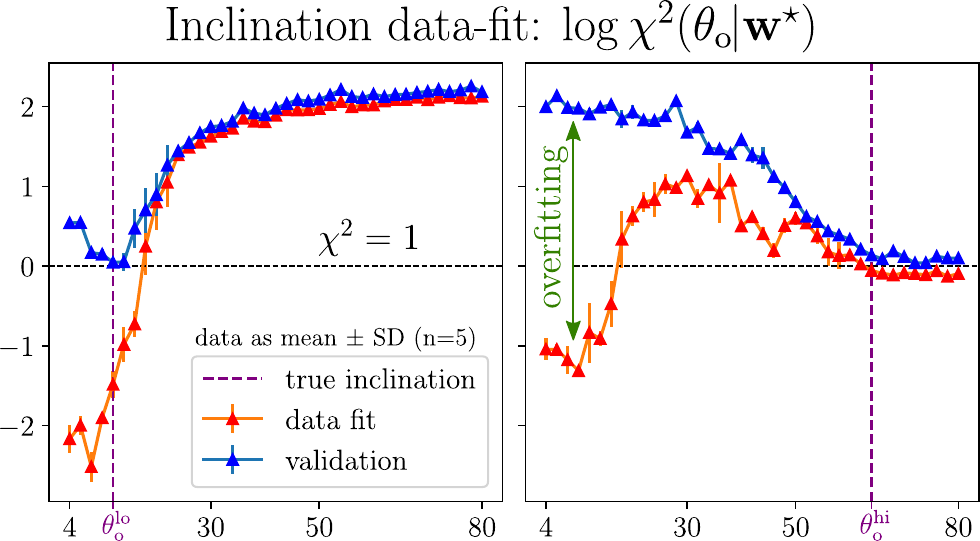}
	\caption{Inclination vs validation $\chi^2$ highlighting the role of the validation $\chi^2$ in model selection. The $\chi^2$ plots are for a {\tt Simple Hotspot} flare with background accretion noise: $\sigma_P=0.01$. The left/right panels show the $\chi^2$ for the true low/high inclinations. The right panel shows how even when the true inclination is $\inchi$, low inclination angles are prone to overfitting the data. The validation $\chi^2$ mitigates over-fitting and enables estimating the underlying inclination angle as the global minimum. For each angle, the 3D recovery is run with five random initializations with error bars indicating the recovery stability.}
	\label{fig:train_val_loss}
\end{figure}
Using the {\tt Simple Hotspot} model with background accretion noise ($\sigma_P$) we show how the conditional $\chi^2(\inc|\netparams^\star)$ enables differentiating $\inclo$ from $\inchi$ (Supplementary Fig.~\ref{fig:train_val_loss}). The right panel of Supplementary Fig.~\ref{fig:train_val_loss} highlights the overfitting gap, illustrating how low inclination angles are prone to overfitting. Intuitively, this is because a high inclination provides a stronger physical constraint, requiring the 3D emission to ``disappear'' behind the black hole for parts of the orbit. Nevertheless, the validation curves mitigate over-fitting and enable estimating $\inc^\star$ (as the global minimum) to within $\pm10^\circ$ of the underlying true inclination. This is because overfit structures do not generalize as well and are less stable under perturbations to pixel positions on the image plane.

\subsection{Sensitivity to background noise}
\begin{figure*}[t]
	\centering \includegraphics[width=\linewidth]{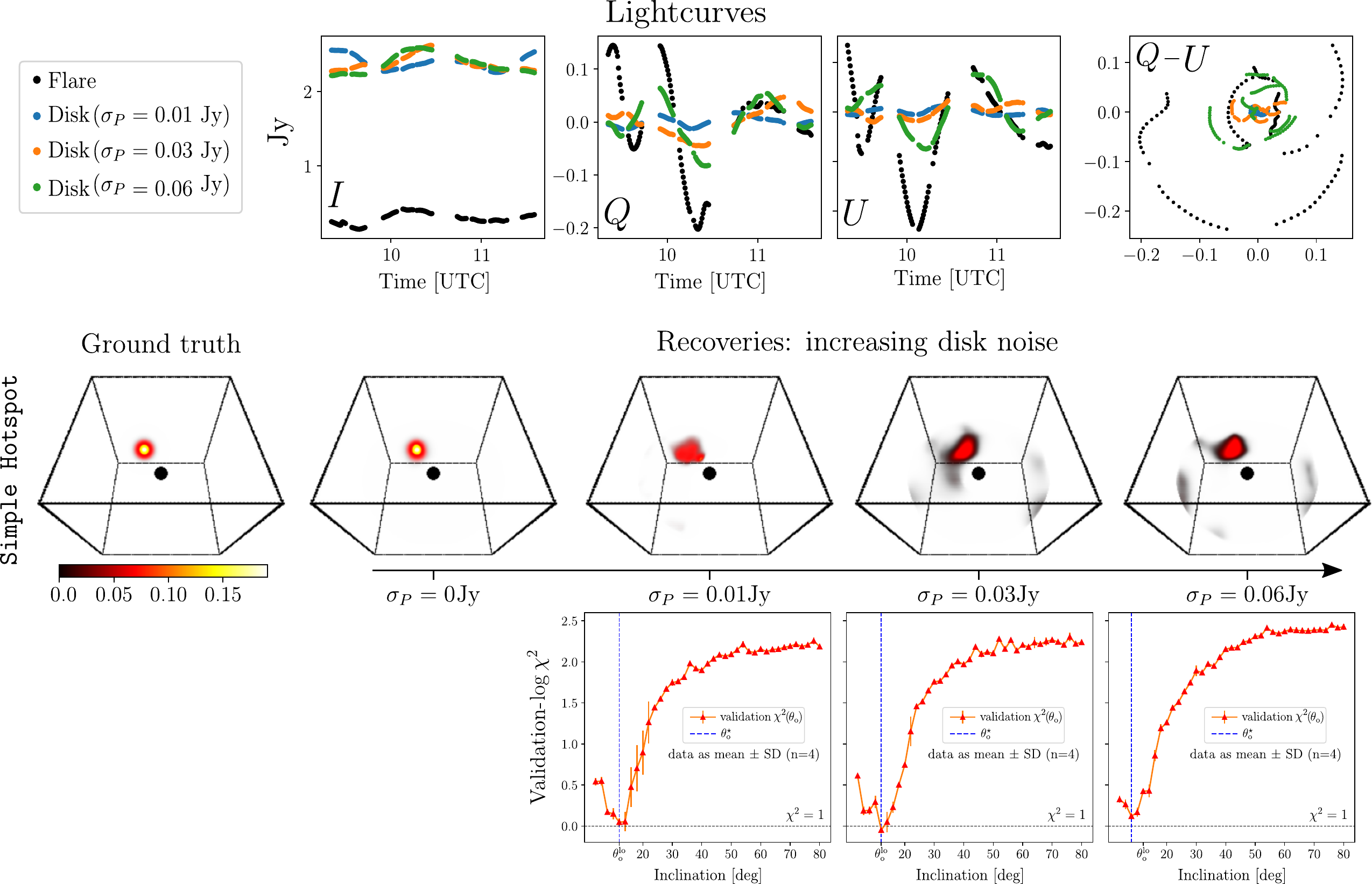}
	\caption{Analysis of the data and reconstructions for three levels of accretion variability. [Top] A comparison between the light curves of the foreground flare and background accretion disk for three levels of disk variability $\sigma_P=\{0.01, 0.03, 0.06\}~{\rm Jy}$. [Middle] 3D reconstructions for the three levels of variability. These volumes represent the true/estimated initial emission at 9.35 UTC. The recovered 3D emission captures the compact underlying true flare structure, size and position. [Bottom] In all cases, the estimated inclination $\inc^\star$ (blue) (estimated as the global minimum) is within $\pm4^\circ$ of the underlying true inclination $\inclo$. The error bars are computed by running the 3D recovery with four random initializations.}
	\label{fig:simple_hs_w_noise}
\end{figure*}

In reality, observed variability is due to both the flare and accretion, however, it is difficult to determine the relative contributions of each source. Any un-modeled accretion dynamics will impact recovery results as our model fully attributes dynamics to orbital motion. We explore this aspect by sampling random fields that mimic an accretion process with varying degree of polarization ($q_f=\{0.25, 0.5, 0.75\}~{\rm Jy}$) producing noise light curves with $\sigma_P=\{0.01, 0.03, 0.06\}~{\rm Jy}$. Similar to the pre-processing of ALMA data~\cite{wielgus2022orbital_supp}, we subtract the time-averaged LP component (the static component of the background emission). A comparison between the simulated {\tt Simple Hotspot} flare and accretion disk light curves, for different levels of variability, is shown in Supplementary Fig.~\ref{fig:simple_hs_w_noise} (top panels). In these simulations, the underlying true inclination is $\inclo$.
For each level of noise, $\sigma_P$, we are able to estimate $\inc^\star$ to within $\pm4^\circ$ of the true value.

using the validation-$\chi^2$ minimum, our approach is able to recover the inclination angle to within $\pm4^\circ$ of the true value (see $\chi^2$ curves at the bottom of Supplementary Fig.~\ref{fig:simple_hs_w_noise}). 
Subsequently, using the estimate $\inc^\star$, we recover the unknown 3D emission. Supplementary Fig.~\ref{fig:simple_hs_w_noise} (middle row) shows a comparison between the recovered 3D emission and the ground truth. As the unaccounted disk variability increases, more emission is attributed sporadically within the volume. Nevertheless, the compact hotspot remains the brightest feature in the recovered 3D volume centered around the true azimuthal position.

\subsection{Recovering different 3D morphology}
\begin{figure*}[t]
    \centering \includegraphics[width=.98\linewidth]{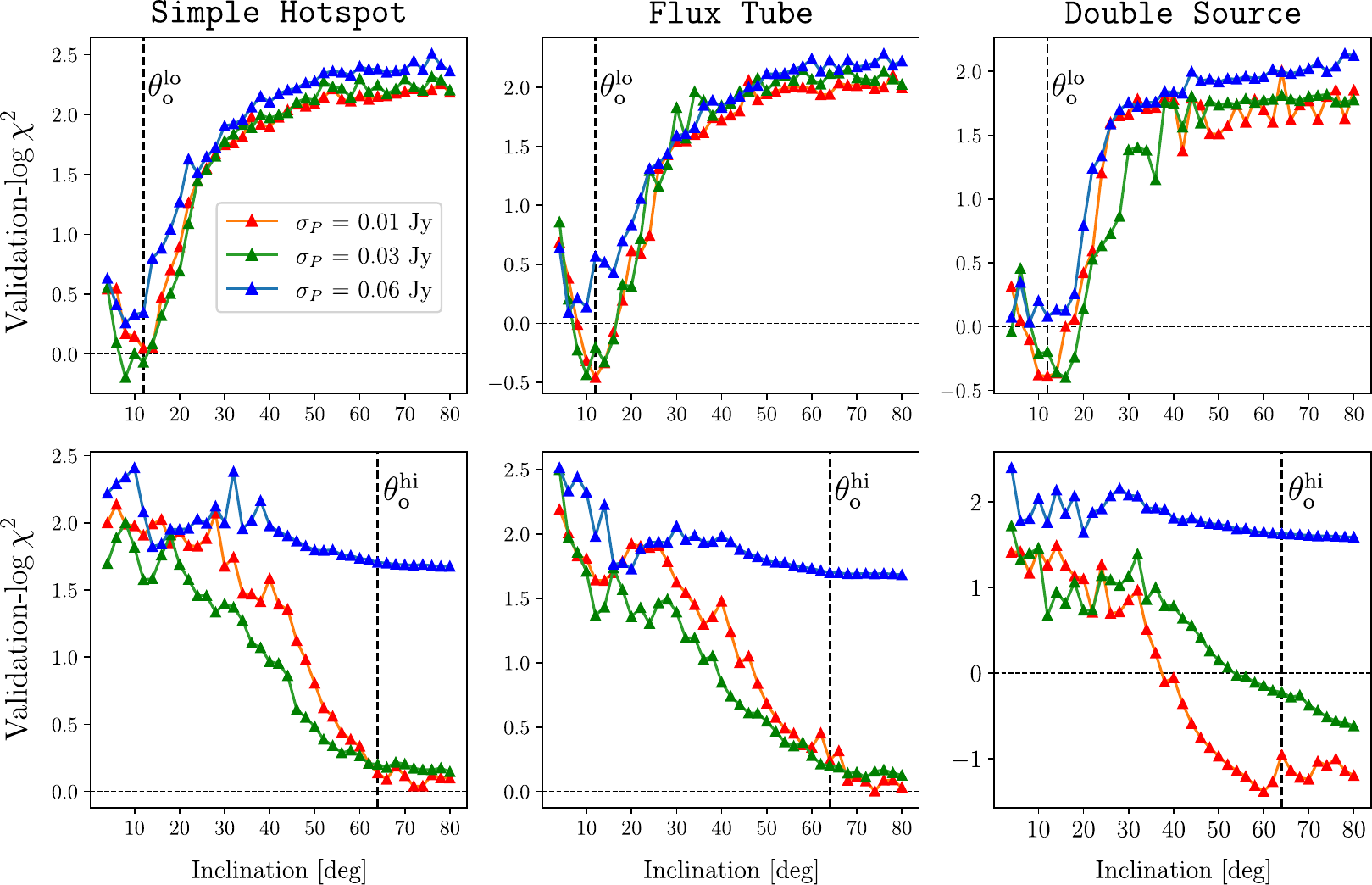}
    \caption{Validation-$\chi^2$ for three synthetic emission structures (Supplementary Fig.~\ref{fig:3d_cases}) under two true inclination angles ($\inclo / \inchi$) and three levels of accretion variability ($\sigma_P$). The red curves correspond to the estimated volumes shown in Supplementary Fig.~\ref{fig:sim_recoveries1}. For $\inclo$ (top row), the minimum is within $\pm6^\circ$ of the true inclination for all noise levels and exactly coincides with the true inclination for low levels of noise (red curves). For $\inchi$ the minimum is within $\pm16^\circ$ of the true inclination for $\sigma_P=0.01, 0.03~\jansky$. While it is difficult to estimate the exact inclination under high accretion variability and inclination angle (bottom row, blue curves), the curve trends enable the detection of high vs low inclinations (compare blue curves at the top and bottom panels).}
    \label{fig:sim_recoveries_incval}
\end{figure*}

\begin{figure}[t]
	\centering \includegraphics[width=\linewidth]{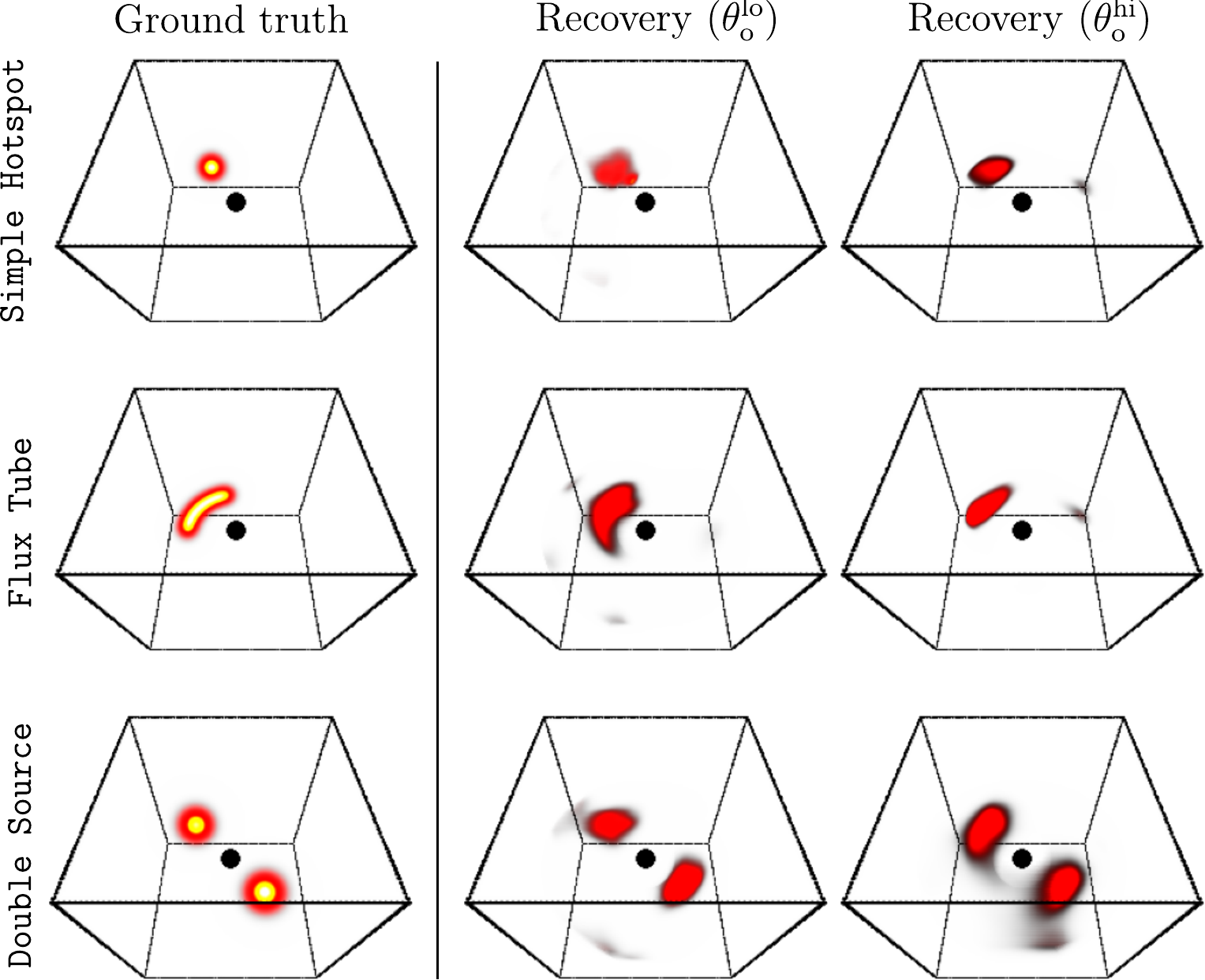}
	\caption{3D recoveries for three simulated structures observed at two (unknown) inclination angles: $\inclo=12^\circ$, $\inchi=64^\circ$. Using synthetically generated light-curves as observations, the 3D reconstructions are able to reconstruct different flare morphologies in the presence of background accretion noise. The recovered structure is given under an inclination $\inc^\star$ that minimizes the validation-$\chi^2$ (Fig.~\ref{fig:sim_recoveries_incval}). Note: this figure is a replica of Fig.~6 in the main text, shown here for convenience.}
	\label{fig:sim_recoveries1}
\end{figure}

In these simulations, we highlight the tomographic ability to recover different underlying emission structures and distinguish between different 3D flare morphologies. 

For each of the datasets we add accretion disk noise with $\sigma_P=0.01~\jansky$ and estimate the inclination angle, $\inc^\star$ as the minimum in the validation-$\chi^2$ (curves are given in Supplementary Fig.~\ref{fig:sim_recoveries_incval}). Our approach is able to estimate $\inclo$ to within $\pm 6^\circ$ for all three levels of accretion noise and 3D fiducial datasets. While low and moderate accretion noise enable detection $\inchi$ to within $\pm 16^\circ$, it is difficult to estimate $\inchi$ under high accretion noise. Nonetheless, even with high accretion noise, the trends of the validation-$\chi^2$ curves enable a robust binary detection of low vs high inclinations.

Supplementary Fig.~\ref{fig:sim_recoveries1} highlights how our approach is able to simultaneously estimate an unknown inclination and 3D structure from unresolved light curve observations. In each case, the recovered 3D captures the unique morphology of the underlying ground truth structure and is smoothed out by the tomographic reconstruction process.

\begin{figure}[t]
	\centering \includegraphics[width=\linewidth]{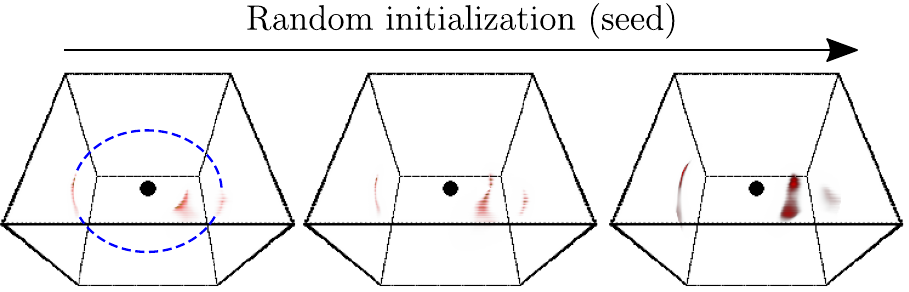}
 \vspace{-0.4cm}
	\caption{3D reconstruction of a stochastic accretion disk (Supplementary Fig.~\ref{fig:accretion_illustration}) without an orbiting bright flare. In these recoveries the observed dynamic signal does not follow the assumptions of orbital tomography. Different initializations (seeds) lead to different local minima. Visualizing the converged solutions gives a sense of the artifacts that would appear in a real data reconstruction (the color bar matches Supplementary Fig.~\ref{fig:simple_hs_w_noise}) The domain boundary is highlighted in blue in the leftmost recovery.}
	\label{fig:inoisy_rec}
\end{figure}

It is instructive to visualize noise present in the 3D reconstructions. To illustrate noisy features we recover the 3D volume of a stochastic disk, without an orbiting bright flare, from light curve observations. In these recoveries the dynamic signal does not conform to the model assumptions made by the orbital tomography algorithm as brightness does not merely orbit and shear but also appears and disappears. Supplementary Fig.~\ref{fig:inoisy_rec} demonstrates the 3D reconstructions with different random seeds (initial conditions for the neural network / 3D volume). Under certain initializations compact emission may appear, however, in contrast to the reconstructions of flares, these components are not stable. The consistent artifacts at the boundary, also present in the real data reconstructions (Figs.~1, 4 of the main text), suggest these are likely an artifact of the reconstruction algorithm.

\subsection{Recovery blind spot}
\begin{figure}[t]
	\centering \includegraphics[width=\linewidth]{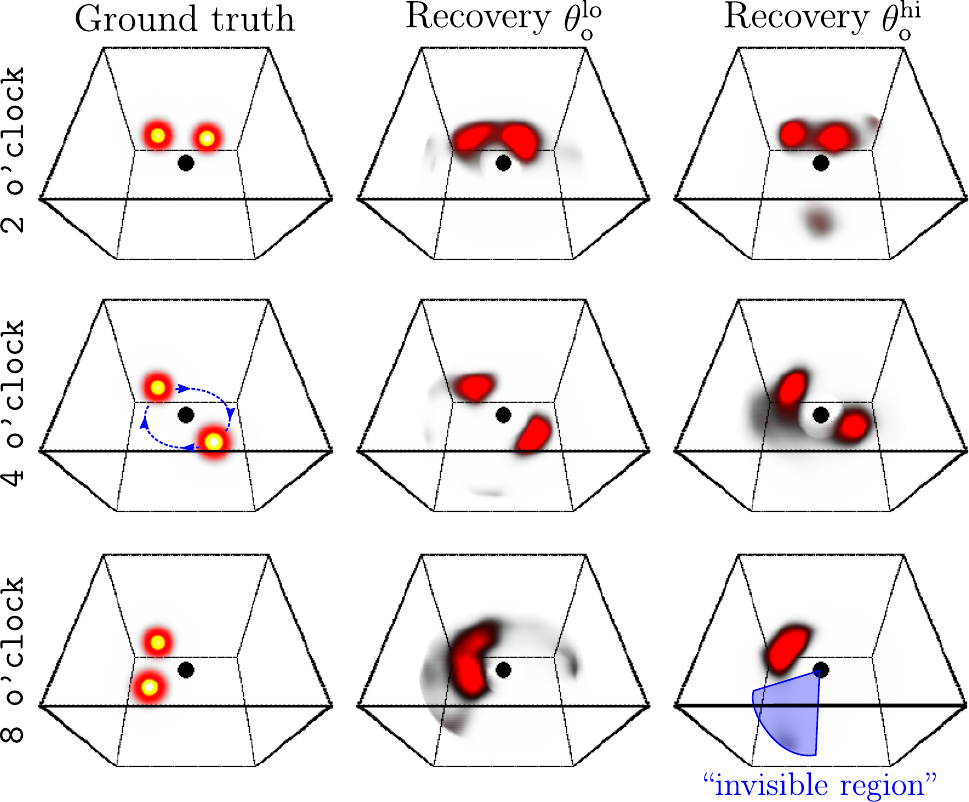}
	\caption{3D recoveries of two bright spots in a clockwise orbit highlight the effects of Doppler boosting. The initial position of one source is fixed {\tt 11 o'clock} and the other at {\tt 2/4/8 o'clock} (different rows). Each column represents recoveries with observations at a different inclination angle $\inclo/ \inchi$. Doppler boosting acts to dim sources moving away from the observer resulting in a blind spot (highlighted in blue) at high inclination angles. Nevertheless, for the {\tt 2/4 o'clock} orbital configurations our approach is able to recover the distinct structure of two emission sources.}
	\label{fig:two_gaussians}
\end{figure}

Close to the event horizon, gas is orbiting at relativistic velocities resulting in a substantial Doppler boosting effect, Doppler boosting effects are particularly strong at high inclination angles where emission moving away appears very faint compared to emission moving towards the observer. This is clearly illustrated in Supplementary Fig.~\ref{fig:3d_cases} where one of the bright spots in the {\tt Double Source} images is extremely faint due to the clockwise rotational motion. For this reason, we anticipate that reconstruction errors will not be distributed equally across the orbital plane. 

To test this hypothesis we simulate observations with two sources, one at a fixed {\tt 11 o'clock} azimuthal position and the other at variable {\tt 2/4/8 o'clock} positions (Supplementary Fig.~\ref{fig:two_gaussians}). The recoveries shown in Supplementary Fig.~\ref{fig:two_gaussians} reveal an ``invisible region'' at high inclination angles where sources are dim due to Doppler boosting. Due to recovery blurring, bright features that are close together merge to form a single region (see {\tt 2/4/8 o'clock} positions in Supplementary Fig.~\ref{fig:two_gaussians}). Nevertheless, these regions contain two distinct brightness peaks attributed to the two hotspots. Doppler boosting is weaker at low inclination angles which are favored by the real data fit.

\section{General Relativistic Ray Tracing}
\label{sec:GR}
This section describes the General Relativistic (GR) equations used to compute geodesic paths and polarized radiative transfer.

\subsection{Kerr metric}
We work in the Kerr black hole spacetime in units where $G=c=1$. A Kerr black hole has mass $M$ and spin $a$. We use Boyer-Lindquist coordinates, where points in the spacetime are indicated by a four-vector $x^\mu=(t,r,\theta,\phi)$.

The structure of spacetime is encoded in the metric $g_{\mu\nu}$, which is a $4\times4$ symmetric matrix. In Boyer-Lindquist coordinates, the nonzero components of the metric are
\begin{align}
g_{tt} &= -(1-2Mr/\Sigma)  \\
g_{rr} &= \Sigma/\Delta \\
g_{\theta\theta} &= \Sigma \\
g_{\phi\phi} &= \Pi\sin^2\theta/\Sigma \\
g_{t\phi} &= g_{\phi t} = -2Mar\sin^2\theta/\Sigma.
\end{align}
where the following are commonly used abbreviations. 
\begin{align}
    \Delta &= r^2 +a^2 - 2Mr \\
    \Sigma &= r^2 + a^2\cos^2\theta \\
    \Pi &= (r^2 +a^2)^2 - a^2\Delta \sin^2\theta \\
    \omega &= 2aMr/\Pi.
\end{align}
The nonzero components of the inverse metric $g^{\mu\nu}$ are
\begin{align}
g^{tt} &= -\Pi/\Delta\Sigma \\
g^{rr} &= \Delta/\Sigma \\
g^{\theta\theta} &= 1/\Sigma \\
g^{\phi\phi} &= (\Delta-a^2\sin^2\theta)/(\Delta\Sigma\sin^2\theta) \\
g^{t\phi} &= g^{\phi t} = -2Mar/\Delta\Sigma.
\end{align}

\subsection{Ray tracing}
We define the image plane of the observer at time $t=0$, at radius $r_o\rightarrow \infty$, and at an inclination angle $\theta_o$ to the angular momentum axis of the black hole (BH). We use Cartesian coordinates $(\alpha,\beta)$ on the image plane, measured in units of $M$. Restoring physical units, the image plane coordinates are in radians with the scale given by $GM/Dc^2$ where $D$ is the distance to the BH. Time is measured in units of $GM/c^3$. The origin $\alpha=\beta=0$ is the line of sight to the black hole at $r=0$, and the $+\beta$ image plane axis is parallel with the BH spin direction (i.e. with the $-\theta$ direction). 

We solve for the trajectories of photons backward from a specific position $(\alpha,\beta)$ on the image plane using a set of differential equations which in Einstein notation can be written as:
\begin{equation}
    \frac{dx^\mu}{d\sigma} = k^\mu.
    \label{eq:ode}
\end{equation}
Here $\sigma$ is the path parameter (``affine time'') and 
\begin{equation}
    x^\mu(\sigma) = \left(t(\sigma), r(\sigma), \theta(\sigma), \phi(\sigma) \right),
\end{equation}
is a parameterized trajectory in spherical (Boyer-Lindquist) coordinates.

Each photon trajectory is defined by three \emph{constants of motion}: $(E,\lambda,\eta)$. $E$ is photon energy at infinity (i.e., observed frequency). The parameters $\lambda$ (angular momentum) and $\eta$ (Carter constant) are fully determined by the photon's coordinates on the observer screen: 
\begin{align}
    \lambda &= -\alpha\sin\theta_o \\
    \eta &= (\alpha^2-a^2)\cos^2\theta_o + \beta^2.
\end{align}
From these constants, at each point along the trajectory, the photon's \emph{covariant momentum} (lower index) $k_\mu$ is given by~\cite{gralla2020null_supp}:
\begin{align}
\label{eq:photonmomentum0}
    k_t &= -E \\
    k_r &= \pm E \sqrt{\mathcal{R}(r)} / \Delta \\
    k_\theta &= \pm E \sqrt{\Theta(\theta)} \\
    k_\phi &= E\lambda,
    \label{eq:photonmomentum1}
\end{align}
where $\mathcal{R}$ and $\Theta$ are \emph{radial and polar potentials}: 
\begin{align}
    \mathcal{R}(r) &= (r^2 + a^2 - a \lambda)^2 - \Delta\left[\eta + (\lambda-a)^2\right]\\
    \Theta(\theta) &= \eta + a^2\cos^2\theta - \lambda^2/\tan^2\theta.
\end{align}
The choice of $\pm$ sign in Eq~(\ref{eq:photonmomentum0}--\ref{eq:photonmomentum1}) depends on the direction of motion (signs reverse at angular and radial turning points when $k^\theta=0$ or $k^r=0$, respectively.). 

Equation \eqref{eq:ode} is described in terms of the {\em contravariant} (upper index) momentum vector $k^\mu$, which is obtained by multiplying the {\em covariant} momentum (Eqs.~\ref{eq:photonmomentum0}--\ref{eq:photonmomentum1}) by the inverse metric
\begin{equation}
k^\mu = g^{\mu\nu}k_\nu.
\end{equation}
To compute geodesic paths $x^\mu(\sigma)$, we solve Eq.~\eqref{eq:ode} using the explicit parametric form described by~\cite{gralla2020null_supp} and implemented in~\cite{kgeo2023_supp}.

\subsection{Fluid Velocity and Redshift}
Outside the ISCO, the four-velocity $u^\mu$ of prograde circular orbits in the equatorial plane is $u^\mu(r)$, where:
\begin{align}
    u^t &= u^t \\
    u^r &= 0 \\
    u^\theta &= 0 \\
    u^\phi &= u^t\Omega,
\end{align}
Since the four-velocity must be normalized so that $u^\mu u_\mu = g_{\mu\nu}u^\mu u^\nu=-1$,
\begin{align}
    u^t &= \frac{1}{\sqrt{-g_{tt} - 2\Omega g_{t\phi} - \Omega^2g_{\phi\phi}}}.
\end{align}

The redshift factor $g$ describes the change in frequency of a photon between the observer's frame at infinity and the frame where it is emitted. 
\begin{equation}
\label{eq:doppler}
    g^{-1} = \frac{\nu_\mathrm{emis}}{\nu_{\rm obs}} = \frac{-k_\mu u^\mu}{E}.
\end{equation}

\subsection{Parallel transport}
Ray-tracing polarization in curved spacetime requires parallel transport of the emitted polarization to the observer screen. We define this operation as a rotation matrix: $\rotation$ (main text: Eq.~5). This matrix multiplies the emission-frame Stokes vector $\bf J$ to rotate the linearly polarized components:
\begin{equation}
\label{eq:paralleltrans}
    \rotation{\bf J} = 
    \begin{bmatrix}
    1 & 0 & 0 & 0 \\
    0 & \cos2\chi & -\sin2\chi & 0 \\
    0 & \sin2\chi & \cos2\chi & 0 \\
    0 & 0 & 0 & 1
    \end{bmatrix}
    \begin{bmatrix}
    J_{I} \\ J_{Q} \\ 0 \\ J_{V}
    \end{bmatrix}.
\end{equation}
The angle $\chi$ at pixel coordinates $(\alpha, \beta)$ is given by~\cite{himwich2020universal_supp}:
\begin{equation}
    e^{2i\chi} = \frac{(\beta+i\mu)(\kappa_1 - i\kappa_2)}{(\beta-i\mu)(\kappa_1 + i\kappa_2)},
\end{equation}
where $\kappa = \kappa_1 + i\kappa_2$ is the complex Penrose-Walker constant, a quantity that is computed in the emission frame and then conserved in parallel transport, and \mbox{$\mu = -(\alpha+a\sin\theta_o)$}.

To compute $\kappa$ in the emitter frame we first convert the coordinate-frame 4-vectors $k^\mu$ and $b^\mu$ for the magnetic field to local 3-vectors $\bf{k}$ and $\bf{B}$ using the tetrad matrices $e^\mu_a$ (which depend on the emission position and fluid velocity) as described in \cite{grtrans_supp} (Equations 36-42). We then compute the local emitted polarization vector $\bf{f}$ from
\begin{equation}
    \mathbf{f} = \frac{\mathbf{k}\times\mathbf{B}}{|\mathbf{k}|}.
\end{equation}
We then convert $\bf{f}$ to a coordinate frame four-vector $f^\mu$ using the transpose tetrad matrix. The constant $\kappa$ is then \cite{grtrans_supp,gelles2021polarized_supp}
\begin{align}
    \kappa &= \kappa_1 + i\kappa_2 = (r-ia\cos\theta)(A-iB)\\
    A&=(k^tf^r-k^rf^t)+a\sin^2\theta(k^rf^\phi-k^\phi f^r)\\
    B&=\left[(r^2+a^2)(k^\phi f^\theta-k^\theta f^\phi)-a(k^tf^\theta - k^\theta f^t)\right]\sin\theta.
\end{align}

\bibliographystyle{plain}

\end{document}